\documentclass[graybox]{svmult}

\usepackage{type1cm}        
\usepackage{makeidx}         
\usepackage{graphicx}        
\usepackage{multicol}        
\usepackage[bottom]{footmisc}

\usepackage{newtxtext}       %
\usepackage{newtxmath}       

\makeindex             

\usepackage[framemethod=TikZ]{mdframed}
\usepackage{lipsum}
\mdfdefinestyle{MyFrame}{%
    linecolor=blue,
    outerlinewidth=2pt,
    roundcorner=20pt,
    innertopmargin=\baselineskip,
    innerbottommargin=\baselineskip,
    innerrightmargin=20pt,
    innerleftmargin=20pt,
    backgroundcolor=gray!20!white}

\mdfdefinestyle{MyFrame2}{%
    linecolor=red,
    outerlinewidth=2pt,
    roundcorner=20pt,
    innertopmargin=\baselineskip,
    innerbottommargin=\baselineskip,
    innerrightmargin=20pt,
    innerleftmargin=20pt,
    backgroundcolor=gray!10!white}    

\usepackage{geometry} 
\usepackage{tikz}

\newcommand{\mybox}[4]{
    \begin{figure}[h]
        \centering
    \begin{tikzpicture}
        \node[anchor=text,text width=\columnwidth-1.2cm, draw, rounded corners, line width=1pt, fill=#3, inner sep=5mm] (big) {\\#4};
        \node[draw, rounded corners, line width=.5pt, fill=#2, anchor=west, xshift=5mm] (small) at (big.north west) {#1};
    \end{tikzpicture}
    \end{figure}
}    

\usepackage[colorlinks=true, allcolors=blue]{hyperref}
\usepackage{hyperref}

\usepackage[style=numeric, sorting=none]{biblatex}
\usepackage{comment}

\usepackage{fancyhdr}
\begin{document}

\title*{NSF-UKRI Bilateral Workshop  \\
Quantum Information Science in Chemistry}
\author{Gregory D Scholes, Alexandra Olaya-Castro, Shaul Mukamel, Adam Kirrander, Kang-Kuen Ni, Gordon Hedley, and  Natia Frank}
%
%

\maketitle
\thispagestyle{plain} 
\vspace{-3.0cm}

\section*{Abstract}
This document summarizes the context and main outcomes of the discussions that took place during the NSF-UKRI bilateral workshop on Quantum Information Science in Chemistry, held on 12-13 February 2024, in Alexandria, Virginia (US). The workshop was jointly funded by the National Science Foundation (NSF) and UK Research and Innovation (UKRI) through the Engineering and Physical Sciences Research Council (EPSRC). It brought together scientific delegations from the United States of America (US) and the United Kingdom (UK). 

\section*{Workshop information}
The workshop was held over 1.5 days during February 12-13, 2024 in Alexandria, Virginia (USA). It was chaired by Gregory D. Scholes and Alexandra Olaya-Castro, and co-organized by teams based in the US and the UK:\\

\textbf{NSF Team:}
\begin{itemize}
    \item Gregorgy D Scholes, Princeton University (\href{mailto:gscholes@princeton.edu}{gscholes@princeton.edu})
    \item Natia Frank, University of Nevada-Reno (\href{mailto:nfrank@unr.edu}{nfrank@unr.edu})
    \item Shaul Mukamel, University of California-Irvine (\href{mailto:smukamel@uci.edu}{smukamel@uci.edu})
    \item Kang-Kuen Ni, Harvard University (\href{mailto:ni@chemistry.harvard.edu}{ni@chemistry.harvard.edu})
\end{itemize}

\textbf{UKRI Team:}
\begin{itemize}
    \item Alexandra Olaya-Castro, University College London (\href{mailto:a.olaya@ucl.ac.uk}{a.olaya@ucl.ac.uk})
    \item Gordon Hedley, University of Glasgow (\href{mailto:Gordon.Hedley@glasgow.ac.uk}{Gordon.Hedley@glasgow.ac.uk})
    \item Adam Kirrander, University of Oxford (\href{mailto:adam.kirrander@chem.ox.ac.uk}{adam.kirrander@chem.ox.ac.uk})
\end{itemize}

Early career researchers  were engaged as “scribes” to document the discussions are: 
\begin{itemize}
    \item Negar Bardaran (Princeton University)
    \item Avery Cirincione-Lynch (Princeton University)
    \item Hallmann Gestsson (University College London)
    \item Ava Hejazi (Princeton University)
    \item Charlie Nation (University College London)
    \item Jane Nelson (Princeton University)
\end{itemize}

The workshop had fifty attendees comprising a diverse group with expertise spanning chemistry, physics and engineering. Including co-organizers and scribes, there were 28 attendees from the US and 22 from the UK.

\pagebreak 

Before the in-person workshop, we distributed a \textbf{“workshop pre-read”} to all participants to get them thinking about the key discussion points:\\

\textbf{Basic concepts in QIS:}
\begin{itemize}
    \item Quantum entanglement
    \begin{itemize}
        \item[] Ryszard Horodecki, Paweł Horodecki, Michał Horodecki, and Karol Horodecki
        \item[] Rev. Mod. Phys. 81, 865 (2009) 
        \item[] DOI: \href{https://doi.org/10.1103/RevModPhys.81.865}{https://doi.org/10.1103/RevModPhys.81.865}
    \end{itemize}
    \item Colloquium: Quantum coherence as a resource
    \begin{itemize}
        \item[] Alexander Streltsov, Gerardo Adesso, and Martin B. Plenio
        \item[] Rev. Mod. Phys. 89,  (2017)
        \item[] DOI: \href{https://doi.org/10.1103/RevModPhys.89.04100 3}{https://doi.org/10.1103/RevModPhys.89.04100 3} 
    \end{itemize}
    \item Quantum resource theories
    \begin{itemize}
        \item[] Eric Chitambar and Gilad Gour
        \item[] Rev. Mod. Phys. 91, 025001 (2019)
        \item[] DOI: \href{https://doi.org/10.1103/RevModPhys.91.025001}{https://doi.org/10.1103/RevModPhys.91.025001} 
    \end{itemize}
    \item Foundations of Quantum Information for Physical Chemistry
    \begin{itemize}
        \item[] Weijun Wu, Gregory D. Scholes
        \item[] arXiv:2311.12238 (2023)
        \item[] DOI: \href{https://doi.org/10.48550/arXiv.2311.12238}{https://doi.org/10.48550/arXiv.2311.12238} 
    \end{itemize}
\end{itemize}

\textbf{Perspectives for QIS in Chemistry:}
\begin{itemize}
    \item Entanglement classifier in chemical reactions
    \begin{itemize}
        \item[] Junxu Li, and Sabre Kais
        \item[] Sci. Adv.5, eaax5283 (2019)
        \item[] DOI: \href{https://doi.org/10.1126/sciadv.aax5283}{https://doi.org/10.1126/sciadv.aax5283} 
    \end{itemize}
    \item Attosecond Physics and Quantum Information Science
    \begin{itemize}
        \item[] M. Lewenstein et al.
        \item[] E-Print Comments: 19 pages, 5 figures, ATTO VIII Conference Proceedings
        \item[] DOI: \href{https://doi.org/10.48550/arXiv.2208.14769}{https://doi.org/10.48550/arXiv.2208.14769} 
    \end{itemize}
\end{itemize}

\textbf{New experiments in QIS with chemical systems:}
\begin{itemize}
    \item Quantum state tomography of molecules by ultrafast diffraction
    \begin{itemize}
        \item[] Zhang, M., Zhang, S., Xiong, Y. et al.
        \item[] Nat Commun 12, 5441 (2021)
        \item[] DOI: \href{https://doi.org/10.1038/s41467-021-25770-6}{https://doi.org/10.1038/s41467-021-25770-6} 
    \end{itemize}
    \item Dipolar spin-exchange and entanglement between molecules in an optical tweezer array
    \begin{itemize}
        \item[] Yicheng Bao et al.
        \item[] Science382,1138-1143 (2023)
        \item[] DOI: \href{https://doi.org/10.1126/science.adf8999}{https://doi.org/10.1126/science.adf8999} 
    \end{itemize}
\end{itemize}

\textbf{Quantum correlations in complex systems:}
\begin{itemize}
    \item Quantum correlations in molecules: from quantum resourcing to chemical bonding
    \begin{itemize}
        \item[] Lexin Ding et al
        \item[] Quantum Sci. Technol. 8 015015 (2023)
        \item[] DOI: \href{https://doi.org/10.1088/2058-9565/aca4ee}{https://doi.org/10.1088/2058-9565/aca4ee} 
    \end{itemize}
    \item Light emission from strongly driven many-body systems
    \begin{itemize}
        \item[] Pizzi, A., Gorlach, A., Rivera, N. et al. 
        \item[] Nat. Phys. 19, 551–561 (2023)
        \item[] DOI: \href{https://doi.org/10.1038/s41567-022-01910-7}{https://doi.org/10.1038/s41567-022-01910-7} 
    \end{itemize}
\end{itemize}


\subsection*{Pre-workshop assessment}

A pre-workshop questionnaire was circulated to assess participants’ views on the opportunities and challenges they envisioned at the intersection between Quantum Information Science and Chemistry. 36 out of the 50 participants responded the questionnaire and the following are key messages conveyed in the responses.
\\

 \textbf{Message 1: There is opportunity of advancement in both directions. }
\begin{itemize}
\item[] \textit{“Quantum chemistry and in particular ultrafast science may allow one to perform operations in a time frame much faster than what is done at the moment.”}

\item[]\textit{“Chemistry also studies big systems, which may be useful for scalability”}

\item[]\textit{“Better understanding of molecular and super-molecular processes.”}

\item[]\textit{“The role of entanglement and interference in determining reaction dynamics and product formation.”}
\end{itemize}

\begin{itemize}
\item[] \textbf{Message 2: There are core scientific challenges to overcome.}
\\
\item[] \textit{“Chemical systems undergo ultrafast decoherence that limits their controllability”}

\item[]\textit{“Showing we are seeing a genuine quantum effect in the first place. Just because we use quantum mechanics to model something doesn’t make it an explicitly QM process”}
\end{itemize}

\textbf{Message 3: We need a shared conceptual understanding of quantum information science.}

\begin{itemize}
\item[]\textit{“By whatever means one has available, using the exponential scaling of the state space of a quantum system to produce beyond-classical capabilities. }

\item[]\textit{“It is linked strongly to the purity of a quantum system”}

\item[]\textit{“Information stored in wave-functions”}

\item[]\textit{“Use quantum coherence and entanglement as a resource”}

\end{itemize}

 \textbf{Message 4: We have key challenges to address in terms of communication and education }

\begin{itemize}
   
\item[] \textit{“I think that the language of QIS is very different from standard physical chemistry (…). The language barrier has been the biggest challenge for my own research group.”}

\item[] \textit{“I think that many chemists face a knowledge gap when it comes to harnessing quantum mechanics.” }

\item[]\textit{“Incoming graduate students in my department are weak in their quantum mechanics knowledge”}
\end{itemize}

 \textbf{Message 5: To innovate we need a variety of resources. }
\begin{itemize}
\item[]\textit{“Funding to try weird science, even if the end goal is not perfectly articulated yet, might be useful in seeding innovation in the area”.}

\item[]\textit{“Form a community of practice” }

\item[]\textit{“Exchange of early career researchers” }

\item[]\textit{“Biannual conference on QIS in Chemistry”}
\end{itemize}

\subsection*{Discussion format}

 The workshop was organised in four discussion-led sessions. Two introductory talks were presented at the beginning of the workshop to set provide a general context from both the QIS and the Chemistry perspectives,  and each discussion-session started with two 15-min thought-provoking talks to motivate the discussion.

 Six discussion tables were set-up. A co-organizer acted as a discussion facilitator, and they were paired up with a scribe. Six to seven participants joined a discussion table. Participants sat at different tables in each discussion session. At the end of a discussion session, one of the members at the table gave a brief 2-minute overview of one most interesting discussion outcome for the table.

 After each discussion session, short interactive quizzes were carried to assess how informative and inspirational the participants found the session talks were. Participants were asked to describe in one word how they found the talks during that session, whether the talks and discussions clarified concepts, and a research question the talks and discussions of the session inspired. Participants gave overall positive feedback to the introductory and the majority of thought-provoking talks. 
\\
The speakers for the introductory talks were: 

\begin{itemize}
    \item[] Gerardo Adesso, University of Nottingham
    \item[] Joseph Zandrozny, The Ohio State University
\end{itemize}

\noindent The speakers for the introductory talks were:
\begin{itemize}
    \item[] Jon Hood, Purdue University  (session 1)
    \item[] Rebecca Ingle, University College London (session 1)
    \item[] Jenny Clark, University of Sheffield (session 2)
    \item[] Ignacio Franco, University of Rochester (session 2)
    \item[] Sabre Kais, Purdue University (session 3)
    \item[] Katherine Inzanie, University of Nottingham (session 3)
    \item[] Sandrine Heuz, Imperial College London (session 4) 
    \item[] Joonho Lee, Harvard University (session 4)
\end{itemize}

\newpage
\section{Introduction: The QIS challenge for chemistry}

Quantum information science (QIS) is a fast-evolving field. One aim is to exploit quantum mechanical phenomena and information sciences to develop technologies with quantum-enhanced functionalities. A second prominent aim is to advance our fundamental understanding of nature at the molecular scale. Concomitantly, researchers across a breadth of fields are interested in seeking new physical, chemical and biological phenomena that are explained by quantum mechanics and which have no classical counterpart. However, a challenge is that important concepts at the core of quantum science can be difficult to appreciate properly. For instance, quantum entanglement, which refers to counter-intuitive correlations among the components of a quantum system, has profound conceptual and technical implications for its identification and understanding in complex molecular systems. Equally challenging is working out of how complex molecular systems can feature in QIS. What are the big questions for chemistry? What are the ways that chemistry can add to and expand the scope of QIS?

The workshop aimed to define and articulate unique “chemistry-centric” opportunities for research directions and open questions at the interface between chemistry and QIS. This is an important distinction, as complex polyatomic molecular systems have both greater degrees of complexity than existing QIS areas, but also greater potential for powerful new outcomes and applications. It is hoped that this report, delivered at the conclusion of the workshop, will provide concrete concepts that suggest how chemistry-oriented research can carve out new and relevant directions for QIS that exploit the strengths and opportunities of molecular systems and the experimental and theoretical tools we have developed in the field. Now that a shift to quantum technologies is becoming evident, a critical mass of researchers are looking for opportunities to impact QIS. The time is right at the this point to educate the wider chemistry community about the kinds of questions and needs that are relevant to QIS. We foresee that influential new research programs that enable chemical science to make an impact will emerge in the coming few years.

The general goal of the workshop was to explore QIS concepts (e.g., entanglement) in quantum states broadly defined (i.e., including non-spin-based quantum states) in molecular systems, also broadly defined. This could include leveraging QIS concepts to advance chemistry research, as well as advancing our understanding of QIS concepts in the context of chemical systems. This workshop also had as a key goal identifying opportunities that leverage capabilities in the US and UK and thus build partnerships.

\section{Background}

A bottleneck for researchers in the chemical science gaining traction in the field of QIS is to understand the subtle concepts that underpin the QIS field. While we have all heard of entanglement, and can appreciate aspects of its nature, the idea is much deeper than meets the eye. Chemistry tends to focus on states, so understanding quantum states is important. Here we provide a brief guide that has been adapted from recent papers \cite{scholes2023, wu2023}. 

A characteristic of quantum systems is that their states, for example energy levels, are quantized: each quantum energy level is associated to eigenstate of the system. We consider an orthonormal basis of these states, so whatever their mathematical form (e.g. they could be $s$ or $p$ atomic orbitals), they can be indicated by state vectors in Hilbert space. For instance, for a two level system, also called a qubit, with an associated basis $\{ |0\rangle , |1\rangle \}$,  a pure quantum state can be  written as $| \psi \rangle = \alpha |0 \rangle + \beta |1 \rangle $, where $\alpha , \beta$ are complex numbers defining amplitude of probabilities. 

On the other hand, observables such as the position operator $\hat{x}$ take a continuous range of values between $-\infty$  and $\infty$, and the eigenstates of $\hat{x}$, which we denote $|x\rangle$, define a continuous variable basis satisfying $\int _{-\infty} ^{\infty} dx\, |x\rangle \langle x | =1$ the closure relationship. In this continuous basis set the state of the system state $|\psi \rangle$ can  then be written as $|\psi \rangle = \int _{-\infty} ^{\infty} dx\, \psi (x)|x\rangle $ and the wavefunction of a system $\psi (x)$ is the probability amplitude at position $x$, i.e. $\langle x |\psi \rangle$. The wavefunction is then sometimes called the  position representation of the state $|\psi \rangle$.

This representation captures the concept that many of the interesting properties explained by quantum mechanics are due to the wavelike nature of the systems. Thus, it is essential to account for phase relationships between the probability amplitudes underlying superpositions and interference.

The quantum mechanical state space needed to encode all the information associated to a compound system with many qubits is constructed from the tensor product of the Hilbert spaces of each qubit in our system, $H = H_{1} \otimes H_{2} \otimes \cdots \otimes H_{n}$. If every qubit were to be definitively set in either the $|0\rangle$ or $|1\rangle$ state, like classical switches, then we obtain $2^{n}$ unique product states in $H$, the same as a classical system of switches. This set of product states forms one possible basis for our general quantum (pure) states that live in the state space $H$. The tensor product structure of the Hilbert space for a compound system has profound implications as the more general states can comprise superpositions of product states. These states are the interesting ones because they have no classical counterpart. 

We are so used to writing quantum wavefunctions as products of basis states that we may not fully appreciate the tensor product basis that we are implicitly using. The main features of the tensor product space are mostly technical. However, a useful intuition is that certain correlations are intrinsically locked into the basis functions of the product structure. For example, the basis state $|0\rangle_A |1\rangle_B$ indicates that the eigenvalues of qubit A and B are complementary, whereas in the basis state $|0\rangle_A |0\rangle_B$ they are identical.

The non-classical properties that make the superpositions of product states special are entanglement and nonlocality. We often discuss ‘pure’ states. Pure states are somewhat unrealistic because they define perfect states in perfect ensembles. Nevertheless, they provide an excellent foundation for understanding what quantum information science offers. The pure states of two entangled qubits, often called the Bell states, are

\begin{equation*}
    |\Psi _{+}\rangle = \frac{1}{\sqrt{2}}\left[ |0\rangle_A |1\rangle_B + |1\rangle_A |0\rangle_B \right]
\end{equation*}
\begin{equation*}
    |\Psi _{-}\rangle = \frac{1}{\sqrt{2}}\left[ |0\rangle_A |1\rangle_B - |1\rangle_A |0\rangle_B \right]
\end{equation*}
\begin{equation*}
    |\Phi _{+}\rangle = \frac{1}{\sqrt{2}}\left[ |0\rangle_A |0\rangle_B + |1\rangle_A |1\rangle_B \right]
\end{equation*}
\begin{equation*}
    |\Phi _{-}\rangle = \frac{1}{\sqrt{2}}\left[ |0\rangle_A |0\rangle_B - |1\rangle_A |1\rangle_B \right]
\end{equation*}
where the subscripts $A,B$ label the qubits. The special feature of these states is they cannot be factored into a state local to $A$ and one local to $B$, as we discuss a little more later in this section. This signals that the states are entangled. A remarkable consequence of entanglement is that a measurement carried out on qubit $A$ instantaneously determines the state of qubit $B$, no matter how far apart the qubits are. This is the principle of nonlocality, challenged in the famous ‘EPR paper’.

It is worthwhile saying a little more about nonlocality to clarify how there is no classical counterpart. It is tempting to make the following classical analogy. Let’s say we have two classical bits $C$ and $D$ that take the eigenvalues red and blue. The bits come in pairs, say joined by a string, and we correlate them by ensuring that if $C$ is red, then $D$ is blue, and vice versa. Now it looks like we have the same outcome as explained above for the entangled quits in $|\Psi _{+}\rangle$ or $|\Psi _{-}\rangle$; that is, we take a pair of bits at random, we look at the color of $C$, say it is blue, then we know that the color of $D$ must be the other color, red in this case, no matter how long the string is.

Entanglement is obviously deeper than this classical correlation. The additional correlations come from the way we make measurements on the state space and it accounts for the quantum mechanical phase, represented by the way the vectors are set up in the combined Hilbert space. Consider a singlet excited state comprising two coupled spins located on separated molecular fragments (like a charge transfer state). The two spins are entangled in the spin eigenstate $|\Psi _{-}\rangle$. We know that if we measure the spin state of the electron on the fragments that they will be opposite. This is the same as the classical correlation we noted in the example of the connected classical bits. The quantum correlation is revealed by noting that the singlet spin state is unchanged by observing it in any reference frame (e.g. the observer can rotate the axes and still measure the same spin-pairing). Bell realized that the extra quantum correlation can often (but not always) be detected by comparing two measurements performed in different reference frames. See Chapter 20 in \cite{bradac2021} for further explanation of this point.

The quantum correlation underlying nonlocality and entanglement has no classical counterpart. It is the key resource for quantum information. Note, however, that there is no ‘extra’ information in a quantum system, it is just that it is encoded differently than in a classical system, which presents opportunities for processing or communicating the information.

Although chemists like to think about wavefunctions, these are not the most general quantum states. In many experiments, a single quantum state in the Hilbert space does allow a complete  description of the system.  The more general quantum state (for a finite dimensional system) is described by a density matrix. The idea of the density matrix is that it represents an element of the set of bounded operators on a Hilbert space $\hat{\rho} \in B(H)$ such that $\hat{\rho}$ is self-adjoint (i.e. Hermitian, $\hat{\rho} ^{*} =\hat{\rho}$), positive, and has unit trace. It enables us to define the expectation value of any operator $\hat{A}$ for the state given by $\hat{\rho}$ via a map $B(H) \rightarrow \mathbb{C}$ ($\mathbb{C}$ is the field of complex numbers) given by
\begin{equation*}
    \phi _{\rho} (\hat{A}) = \text{Tr}(\hat{\rho} \hat{A}) = \text{Tr}(\hat{A} \hat{\rho} ) .
\end{equation*}

The density matrix is built up by an average over the relevant states $|\Psi \rangle$ in our Hilbert space, $\hat{\rho} = \langle |\Psi \rangle \langle \Psi | \rangle $, where $\langle \cdots \rangle$ means average. We can immediately establish that $\hat{\rho}$ is independent of the (unknown) absolute phase of $|\Psi \rangle$. For example, say that $|\Psi \rangle = \exp(i\alpha) |\Phi \rangle$, so that $|\Psi \rangle$ and $|\Phi \rangle$ differ only by overall phase. But notice that $|\Psi \rangle \langle \Psi | = |\Phi \rangle \langle \Phi |$.

Information about whether or not our system is entangled, or contains other quantum correlations, is encoded in the state. However, working out if a state is entangled or not is a difficult problem, and exponentially difficult as the number of basis states increases. It turns out to be most clear if we first explain when a state is not entangled—that is, when it is separable. A state is separable in the Hilbert space comprising the tensor product of the Hilbert spaces of the subsystems $A$ and $B$, $H = H_{A} \otimes H_{B}$ if it can be written 
\begin{equation*}
    \hat{\rho} = \sum _{i=1}^{N}p_{i}\hat{\rho} _{i}^{A}\otimes \hat{\rho} _{i}^{B} .
\end{equation*}

If two subsystems, $A$ and $B$, are entangled, then we cannot construct the composite state using only what we know about the subsystems separately, unlike separable states. This is because the superpositions of product states introduce non-additive correlations into the quantum probability distribution that are encoded in the density matrix of the composite system, but hidden to measurements performed on each subsystem separately.

The first example is very simple. We perform a sequence of measurements on the product state of our composite system comprising the qubits A and B. We find that 25\% of the time we measure $|0\rangle_A |0\rangle_B$, 25\% of the time we measure $|1\rangle_A |0\rangle_B$, 25\% of the time we measure $|0\rangle_A |1\rangle_B$, and 25\% of the time we measure $|1\rangle_A |1\rangle_B$. The density matrix is diagonal, with values of $\rho _{ii} = 1/4$. We write it with rows and columns indexed as $|0\rangle_A |0\rangle_B$, $|0\rangle_A |1\rangle_B$, $|1\rangle_A |0\rangle_B$, $|1\rangle_A |1\rangle_B$:
\begin{equation*}
    \hat{\rho} _{AB} = \hat{\rho} _{A} \otimes \hat{\rho} _{B} = \begin{pmatrix}
        \frac{1}{4} & 0 & 0 & 0 \\
        0 & \frac{1}{4} & 0 & 0 \\
        0 & 0 & \frac{1}{4} & 0 \\
        0 & 0 & 0 & \frac{1}{4} \\
    \end{pmatrix} .
\end{equation*}

\noindent This is a fully mixed (classical) state which, of course, is separable.

It is not so straightforward to display more general examples. Peres [19] devised a clever technique to write a guaranteed mixed state, and this provides a good model for us. Consider a pure singlet state ($|\Psi _{-}\rangle$). This singlet state will make up a fraction $x$ of our mixed state, while the remaining $(1 - x)$ fraction is a ‘random fraction’, comprising equal admixture of the singlet and three triplet states to produce a fully mixed fraction of the state. The random fraction contributes a value of $(1 - x)/4$ to each diagonal element of $\hat{\rho} _{\text{mixed}}$:

\begin{equation*}
    \hat{\rho} _{\text{mixed}} = \begin{pmatrix}
        \frac{(1-x)}{4} & 0 & 0 & 0 \\
        0 & \frac{(1+x)}{4} & -\frac{x}{2} & 0 \\
        0 & -\frac{x}{2} & \frac{(1+x)}{4} & 0 \\
        0 & 0 & 0 & \frac{(1-x)}{4} \\
    \end{pmatrix} .
\end{equation*}

If we consider $\hat{\rho} _{\text{mixed}}$ with $x=0$, then the state is fully mixed, by design, and the density matrix is diagonal, like the previous example. If $x=1$ the state should not be separable. It can be shown (see appendix A in \cite{scholes2023}), this state is separable only if $x< 1/3$, defining a cross-over between these extreme cases of separability and non-separability.

\section{Concepts}

The characterization of the differences between classical and quantum states has transformed our view of the atomic and molecular components of matter. Quantum information science has shown that such differences not only have a fundamental role in our understanding of nature and the boundaries between classical and quantum systems, but that such differences can be exploited as a physical resource in the development of technologies with either enhanced or fully new capabilities for a variety of tasks, such as information transfer, sensing, communication, computation, and metrology. 

Here we summarize some of the concepts that are important for characterizing the quantum resources associated to a quantum state. 

\subsection{Quantum resource theories}

Quantum resource theories are important mathematical and conceptual frameworks that allow the quantification of specific quantum phenomena such us coherence, entanglement, nonclassicality, quantum thermodynamics, or asymmetry, and formally investigate their usefulness to perform a given task.  Resource-theoretic frameworks divide quantum states and operations into two sets, free and resource states, and provide the tools to characterize the operational significance of the latter \cite{chitambar2019}. Resource states cannot be generated via the corresponding free quantum operations. The quantum resource theory then gives insight into the tasks that are possible within this restricted set of operations. 

For example, in the resource theory of thermodynamics \cite{lostaglio2019}, free states are thermal equilibrium states defined by the inverse temperature and the accompanying free operations are thermal operations (i.e. interactions with a large bath at thermal equilibrium). Thermodynamic transformations must satisfy a family of constraints, known as a family of second laws. 

The synergies between the conceptual framework of quantum resource theories and chemistry are beginning to be explored \cite{yunger2020}. The application of resource theories in photochemical scenarios can expand our understanding of upper and lower bounds of photoisomerization yield in molecular switches besides providing a new view of such systems as quantum clocks \cite{yunger2020}

An open question in this field is a resource theory of quantum dynamics. Developments in this area have the potential to impact our understanding of quantum effects in photochemical processes. 

\subsection{Coherence}

Quantum coherence of the state of a system is arguably the most fundamental quantum feature of single quantum systems that tells us whether measurements of observables are compatible with the system being in a superposition of orthogonal states with respect to a reference basis. Incoherent states are therefore those with a diagonal density matrix in the selected basis, while incoherent operations (which can be defined in a variety of ways) do not create or increase coherence \cite{baumgratz2014, streltsov2017}.

Coherence as a quantum resource is often quantified by distance-based measures which indicate ‘how far’ is a given quantum state from the closest incoherent (diagonal) state in a chosen basis \cite{streltsov2017}.  A proper measure of coherence satisfies key conditions such as being monotonous under incoherent operations. The relative entropy of coherence and the $L_1$- norm of coherence are useful quantifiers of coherence as they admit closed expressions and have physically intuitive interpretations. 

The basis-dependence of coherence introduces an ambiguity: an incoherent state in a selected basis can become coherent under a basis transformation. To address such ambiguity studies have introduced the concept of basis-independent coherence \cite{ma2019} where the incoherent state for reference is the maximally mixed state which is a state that remains incoherent under a basis transformation.

\subsection{Entanglement}

Quantum entanglement refers to unique non-local correlations among the subsystems of a bipartite or multipartite quantum system. As eloquently presented in \cite{horodecki2009}, a key conceptual aspect of entanglement is, in Schr\"odinger words, the fact that “the best possible knowledge of a whole does not include the best possible knowledge of its parts.” This means that an entangled quantum state of a composite system contains more information than all the individual states of the parts.  

Entanglement itself cannot be used to transmit information but the violations of Bell-type inequalities associated to entanglement are key for encoding information in ways that are not possible by classical means. 

Quantification of entanglement is well understood for bipartite systems \cite{horodecki2009}. Examples of such relevant quantifiers are the concurrence (which for a pair of qubits gives a simple measure for the entanglement of formation), and the entropy of entanglement quantified by Rényi entropies  (particularly relevant for many-body systems). For pure states, this latter refers to the information that can be obtained by doing partial measurements on a subsystem while ignoring information about (tracing over the state of) the remaining part. Such partial measurements result in a statistical mixture with a degree of mixedness that depends on the amount of bipartite entanglement.  

While bipartite entanglement is well understood, the characterization and quantification of multipartite quantum correlations is very much an open challenging problem \cite{frerot2023}. 

To gain a deeper understanding of entanglement it is useful to understand separable states which are states that can be prepared by distant laboratories using only local operations and classical communication.  Separable states are described as tensor product states of the density matrices of the individual subsystems or any statistical mixture of such product states. Entangled states are therefore non-separable.  Separability of bipartite states is well understood for low dimensional systems as in that case there is the well-known Peres-Horodecki criterion based on positivity of partial transposition. However, for compound systems of many particles or higher dimensionality there is no universal separability condition. In that case the notion of entanglement witness becomes important. 

Entanglement witness for operators detecting entanglement refers to an inequality for the expected value of such operator which is valid for all separable states but that is violated if and only if the state is entangled i.e. $\text{Tr}[\hat{W}\hat{\rho} _{separable}]>0$ while $\text{Tr}[\hat{W}\hat{\rho} _{entangled}<0$, where $W$ is a Hermitian operator\cite{terhal2001}.  The most attractive feature of detecting entanglement via an entanglement witness is that there is no need for a full reconstruction of the quantum state (and therefore no need for quantum state tomography) as it relies solely on measurements of a mean value of the specific observable.  Entanglement witnesses are  therefore very relevant from an experimental perspective.

\subsection{Quantum correlations beyond entanglement}

Exploiting the quantum nature of chemical systems requires a discrimination of the classical and quantum correlations in a polyatomic system.   It is known that the quantum correlations among subsystems of a multipartite compound  transcend entanglement and one important quantifier of such general quantum correlations is quantum discord \cite{henderson2001, ollivier2001}.  

To understand the concept of quantum discord one starts by considering a joint bipartite system AB with associated density matrix $\hat{\rho} _{AB}$, and focus on the total shared information between subsystems $A$ and $B$ which is quantified by the quantum mutual information defined as $I(A : B) = S(\hat{\rho} _{A}) + S(\hat{\rho} _{B}) - S(\hat{\rho} _{AB})$, where $S(\hat{\rho})=-\text{Tr}[\hat{\rho}\, \text{ln}\hat{\rho}]$ is the von Neumann entropy and the reduced density matrices $\hat{\rho} _{A} = \text{Tr}_{B}[\hat{\rho} _{AB}]$, $\hat{\rho} _{B} = \text{Tr}_{A}[\hat{\rho} _{AB}]$ are the reduced density matrices for subsystems $A$ and $B$, respectively. The goal is then to decompose the shared quantum information into classical correlations and quantum correlations.  This problem was first addressed by Henderson and Vedral \cite{henderson2001} and then independently by Ollivier and Zurek \cite{ollivier2001}

The classical correlations between $B$ and $A$, which we label $J(B|A)$, are the difference in von Neumann entropy of subsystem $B$ before and after a particular measurement is acted on subsystem $A$, i=.e.,
\begin{equation*}
  J(B|A) = \max _{A^{\dagger}_{i}A_{i}} \left\lbrace S(\hat{\rho} _{B}) - \sum _{i}p_{i}S(\hat{\rho} _{B} ^{i}) \right\rbrace  ,
\end{equation*}
where $\rho _{B} ^{i} = \text{Tr}[\hat{A}^{\dagger}_{i}\hat{A}_{i}\hat{\rho} _{AB}]/p_{i}$ is the residual state of $B$ after measurement of $\hat{A}^{\dagger}_{i}\hat{A}_{i}$ (positive operator valued measurements) on subsystem $A$ and $p_{i}=\text{Tr}_{AB}[\hat{A}^{\dagger}_{i}\hat{A}_{i}\hat{\rho} _{AB}]$ is the probability of this outcome.  Note $J(B|A)$ is asymmetric and therefore it will be different for the classical correlations from subsystem $A$ to $B$, which we would label $J(A|B)$.   

The portion of correlations of the mutual information that is not classical must therefore be quantum, and is known as the quantum discord i.e. $D(B|A)= I(A : B) - J(B|A)$. This discord quantifies the amount of information that can be never classically extracted via a local measurement on $A$.  

One of the most interesting aspects of the concept of discord is that separability of the density matrix describing a pair of systems does not guarantee a vanishing discord. This means that for mixed states discord can arise even when entanglement vanishes. For pure states, entanglement and discord are equivalent. 

The applications of these concepts in physical chemistry scenarios, for example, for the the characterization of electronic structure, is providing a new view on chemical boding in molecular systems \cite{ding2022}.

\subsection{Nonclassicality of continuous-variable states }

The quantum optics field has developed a solid framework to investigate quantum properties of radiation field states. In this context, the ‘classical’ states are the Glauber-Sudarshan coherent states $|\alpha \rangle$ which are the right eigenstates of the field annihilation operator $\hat{a}$, i.e. $\hat{a}|\alpha \rangle = \alpha |\alpha \rangle$ with $\alpha$ being a complex number. Quantum behaviour with no classical counterpart, namely,  nonclassicality, arises if the state of the field cannot be expressed as a statistical mixture of such classical coherent states defining a valid probability measure. This then leads to non-positive values of the phase–space quasi-probability representation of the field state, such as negativities in the Glauber–Sudarshan $P$-function or its regularised version.  

Recent works \cite{kwon2019} have unified the concept of nonclassicality investigated via the negativity of the Glauber–Sudarshan $P$-function and the concept of coherence as a resource.  It has been shown that in certain estimation tasks photon states that lead to a negativity in the $P$ function provide metrological enhancement over all classical states. This metrological power is found to be a measure of nonclassicality based on a quantum resource theory that does not increase under linear optics transformations. 

Investigations of quantum correlation in photon states have also suggested that there are quantum correlations that are neither entanglement nor discord which can be accessed via negativities in phase-space for statistical mixtures of photon-number states \cite{ferraro2012, kohnke2021}.

\section{Specific Aims of the Workshop}

The workshop enabled a diverse group of researchers to brainstorm, flesh out, and define the following aims:

\begin{itemize}
    \item[(a)] Identify concretely how QIS matters in chemistry;
    \item[(b)] Articulate some of the most pressing and interesting research questions at the interface between chemistry and QIS; “chemistry-centric” research question relevant to QIS;
    \item[(c)] Propose in what ways and new directions the field should innovate, in particular where a chemical perspective is essential;
    \item[(d)] Explain examples of recent research in chemistry that inspire scrutiny from a QIS perspective.
\end{itemize}

To address these aims, the workshop was structured around four discussion topics:    

\begin{enumerate}
    \item New experiments enabled by QIS or tools that give insight into QIS phenomena
    \item Chemistry that is governed by or optimized by QIS and opportunities presented by chemical systems as platforms for studying QIS
    \item Fundamental studies of quantum correlations enabled by chemical/molecular systems
    \item Measurement of quantum correlations and non-classical properties of complex systems. If we are going to look for non-classical properties, we need to be able to detect them.
\end{enumerate}

\section{Outcomes of the Discussion}

The discussion covered many angles and aspects of the questions. In order to give a flavor of the rich flow of ideas, we describe some of the topics discussed at the tables. The executive summary gives an overview and interpretation of the questions and some more detailed suggestions for answers.

\subsection{New experiments enabled by QIS or tools that give insight into QIS phenomena}

\subsubsection{Executive summary}

Incisive experiments are key to discovering how quantum information can be leveraged in chemical systems, for revealing the presence of quantum effects or function. In the course of future research, it is likely that there will be relevant further developments in entangled photon spectroscopy, quantum imaging, quantum sensing. 

\begin{figure}[!ht]
\centering
\includegraphics[width=1.0\linewidth]{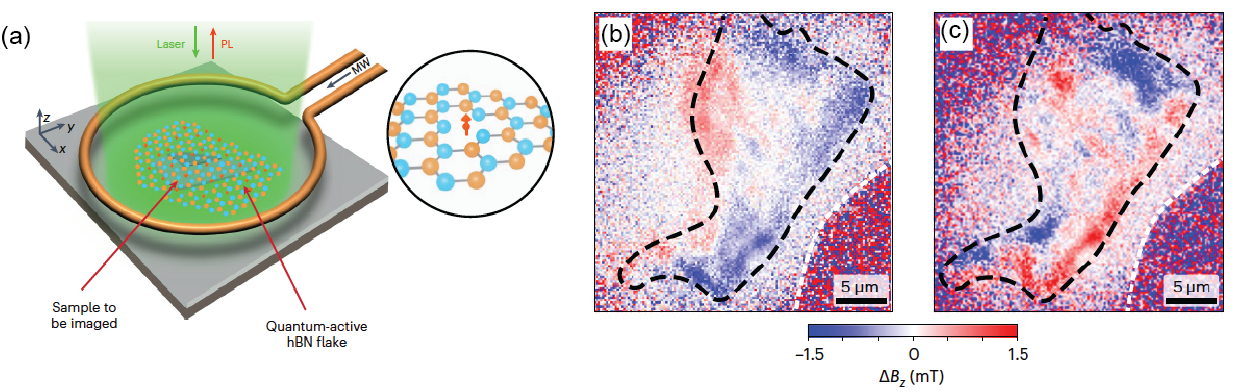}
\caption{\label{fig:1r}Quantum enhance imaging. See text for explanation. Reproduced from \cite{healey2023}}
\end{figure}

Enhancing imaging using quantum science has been gaining ground over recent years \cite{healey2023, bradac2021, kianinia2018}. One example is shown in Fig \ref{fig:1r}. The key breakthrough was to discover two-dimensional materials that can host single spins in defect sites, like those defect centers well-known to exist in diamond, for example. Hexagonal boron nitride (hBN) has been found as an effect two-dimensional substrate that hosts such defect. Fig \ref{fig:1r}a shows a flake of hBN containing single spin defects (red arrows). The sample to imaged is placed on top of the flake and the defect spin states are controlled with the aid of a microwave field \cite{healey2023}. These single spin emitters can thus be used as a way to map the magnetic properties of a material imaged on top of the two-dimensional substrate using spatially-resolved optically-detected magnetic resonance (ODMR). An example is shown in Fig \ref{fig:1r}b, where a CrTe$_2$ ferromagnet is imaged—a map of the stray magnetic field emanating from the CrTe$_2$ flake is shown. In Fig \ref{fig:1r}c the direction of the bias magnetic field is reversed, showing a partial magnetic field reversal. In addition to further advances in the area of quantum-enhanced imaging, these advances have, and will continue to, stimulate a range of studies of the photophysics and defect spin states in two-dimensional materials. 

Effort is needed to develop new experiments based on methods including ultrafast spectroscopy, time-resolved x-ray spectroscopy and scattering, and spectroscopy with squeezed light that detect mechanistic detail relevant to QIS. Oftentimes spectroscopic measurements focus on dynamics, then quantum phenomena are inferred by theoretical models for the data. This can be a powerful approach when the spectroscopic measurement uncovers signatures of mechanism that defy alternative explanations.  It may, however, be more impactful for the QIS field, if the experiments themselves demonstrate the existence of quantum phenomena. We suggest this as a major challenge for the field.

A road block for conceiving new experiments, like those proposed above, can be the focus on the density matrix. The density matrix characterizes a quantum state, and its evolution, and has been a center-point of QIS. However, the complexity of molecular systems makes measurement of the state intractable. New kinds of experiments that reveal glimpses of properties of the state will be of great importance. Experiments that enable unitary transformations of the state space are likely out of reach, but would be transformative because they would enable comparative measurements in different bases. Another rich avenue for experiments would be to study projective measurements and the outcome of those measurements on the state space—which does not necessarily “collapse” completely, but is changed by the measurement \cite{Ozawa1998, JacobsSteck}.

An example shows that ultrafast spectroscopic experiments can reveal some information about the density matrix of the system. While this idea is embedded in the double-sided Feynman diagram “Rosetta Stone” formalism for understanding nonlinear spectroscopy \cite{mukamel1995}, it was especially highlighted by the observations and analysis of coherent oscillations seen in two-dimensional electronic spectroscopic studies of photosynthetic light-harvesting complexes \cite{engel2007, collini2010, fleming2024temp}. These oscillations were interpreted to reflect dynamics of the off-diagonal elements of the system’s density matrix. Measuring dynamics in the off-diagonal part of the density matrix can, in turn, unlock the possibility of detecting and characterizing quantum coherence—so this was a very exciting development. The reason that 2DES can shed light on the ultrafast time evolution of the density matrix is explained by the way the signal is generated systematically from a sequence of impulsive perturbations of the density matrix of the system \cite{mukamel1995,cho2005, branczyk2014} 

\begin{figure}[!ht]
\centering
\includegraphics[width=1.0\linewidth]{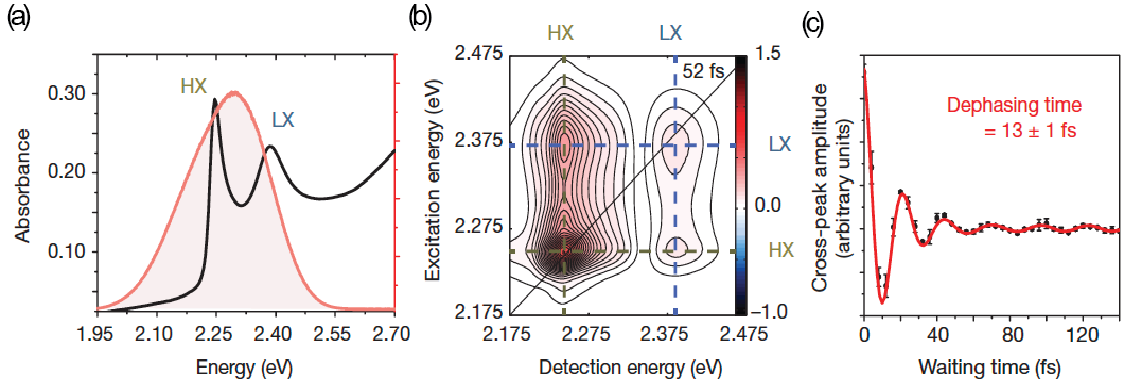}
\caption{\label{fig:2r}(a) Absorption spectra of CdSe nanoplatelets (black line) showing the HX and LX exciton transitions. The spectrum of the laser pulses used in the 2D spectroscopy experiments is shaded orange. (b) 2D electronic spectrum recorded at a pump–probe delay time of 52 fs. (c) Amplitude oscillations in the lower cross-peak of the rephasing 2D spectrum for a CdSe/CdZnS nanoplatelet (the real part with population relaxation subtracted) as a function of the waiting time. Reproduced from \cite{cassette2015}.}
\end{figure}

In Fig. \ref{fig:2r} we show an example of how electronic coherence is detected using two-dimensional electronic spectroscopy (2DES). The study was performed \cite{cassette2015} at ambient temperature on a solution of CdSe nanoplatelets, that show a characteristic pair of absorption peaks labeled HX and LX (the heavy hole and light hole excitons). A short laser pulse excitation sequence excites both these states in superposition, and the experiment resolves the subsequent evolution of the superposition state. It achieves that by detecting the way the off-diagonal terms in the density matrix evolve over time—specifically, they oscillate at the frequency difference of the HX and LX resonances. This principle is easily demonstrated by solving the time-dependent Schrödinger equation for coupled two-level systems. There are techniques that can be exploited to see more clearly how the density matrix is mapped on the 2DES signal \cite{branczyk2014, turner2012}. Here we show only the total signal, Fig \ref{fig:2r}b. The oscillations in one of the cross-peaks are shown in Fig. \ref{fig:2r}c. Notice that the coherence decays quickly, caused mostly by inhomogeneous dephasing of the ensemble. 2DES clearly has the potential to give relevant insights into the system’s density matrix indirectly, but resolving the density matrix with more incisive detail remains an open question.

It is important to note that ultrafast spectroscopy tends to focus on the single-excitation subspace of the full tensor product state space. Doubly-excited states and so on can be detected, but they are very short-lived. This suggests that such experiments are often better designed to investigate a subset of quantum states, a class of coherent states. In the case of excitonic systems, coherent states can be identified with delocalized exciton states. Coherence can serve as a resource for quantum information, and this area of study is highly relevant for chemical systems. In Box 1 we define coherence and the resource theory for coherence. 

A central challenge identified by the workshop is the need to develop structure-function relations that can guide the design and operation of molecular-based quantum devices. In particular, we need to identify robust molecular design principles that can be used to generate quantum subspaces with protected quantum coherences or that offer controllable entanglement. Systematic progress clearly requires developing experimental and theoretical methods to quantify entanglement and coherence, but also an understanding of how the environment influences the dynamics of the system~\cite{gustin2023mapping, kim2024general} and how this influence can be modulated via chemical design\cite{hart2021engineering}. For example, femtosecond resonance Raman experiments can now be used to reconstruct spectral densities that quantify the interaction between electronic transitions with their vibrational and solvent environment. These spectral densities can be used to map how coherence is lost~\cite{gustin2023mapping} and energy dissipated~\cite{kim2024general} by the molecular environment and identify which component of the molecular environment is the most responsible for the decoherence.  This information can then inform efforts in molecular design that seek to take advantage or mitigate the detrimental effects of decoherence.

\begin{figure}[!ht]
\centering
\includegraphics[width=1.0\linewidth]{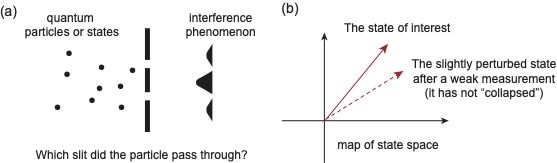}
\caption{\label{fig:wm}Weak measurement illustration (a) Canonical example of quantum interference, where observing the particles as they pass through one of the slits collapses the state and suppressing interference.   (b) In weak measurement, we perform a measurement that gives only partial information about, for example, which slit the particle passed through. That measurement perturbs the state, but does not collapse it. We can build up information on the pathway and the interference phenomenon by accumulating information from many measurements.}
\end{figure}

One area that is underdeveloped, certainly with respect to molecular systems, is weak measurement. Weak measurement and closely related protocols use the idea that many measurements that each very weakly perturb the system can be used to reconstruct information about a quantum state (or its evolution) that is typically hidden by the wavefunction “collapse” when a measuring device indicates a classical reading \cite{aharonov2009}. These weak measurements have also been called ‘protective measurements’ \cite{aharonov1993}. Fuchs and Peres \cite{fuchs1996} discuss the trade-off between information gain and disturbance to the system, Fig. \ref{fig:wm}. Weak measurement has been applied to problems in quantum optics but not yet chemical systems. There might be interesting ways to exploit a weak measurement protocol in spectroscopy, for instance. It might provide a new way to observe quantum mechanical evolution in ultrafast chemical dynamics.

\begin{mdframed}[style=MyFrame, nobreak=true]
\begin{Large}
    \textbf{Box 1: Coherence}\\
\end{Large}

In \cite{scholes2017} we defined coherence as follows: “Coherence can be classical or quantum mechanical and comes from well-defined phase and amplitude relations where correlations are preserved over separations in space or time.” In classical scenarios, waves and their interference come to mind. In QIS we can define quantum coherence as indicated by off-diagonal elements of the density matrix. It is, therefore, evidently basis depend.  The kind of coherence that is discussed in the present work is coherent delocalization. In the site basis a perfectly delocalized exciton state has off-diagonal elements in the density matrix. However, if we choose the eigenstate basis, the density matrix is diagonal. 

Relevant to function related to QIS, we should consider how coherent states can serve as quantum resources \cite{branczyk2014, cassette2015, turner2012, scholes2017, oevering1987, paddon1994, olshansky2019}. According to quantum resource theories \cite{streltsov2017}, the ingredients needed for quantum operations are free states, free operations, and resources. Free states are generally states that are readily available. Similarly, free operations, which map free states to free states, are easily implementable. The crucial ingredients are resources. One systematic framework for quantifying how coherence can serve as a resource is described by Baumgratz, Cramer, and Plenio \cite{baumgratz2014}.

Coherent states likely have attributes that could be used for functions relevant to QIS in chemical systems by making use of interference phenomena or other non-classical correlations. Potential applications include sensors, detectors, switches, communication systems, or transport of excitons and charges. 
\end{mdframed}

\subsubsection{Bullet point summary of the discussion}

\begin{itemize}
    \item How can we realistically extract information from experiments that allows partial reconstruction of the density matrix?
    \item How should we decide how to partition systems into the system of interest and the environment (bath states)?
    \item Most chemical measurements are averages, whether an ensemble average over many copies of the system or a time average involving multiple measurements of a single molecule. Thus the environment is very important because of its random influence during the accumulation of an average. Studying averaging in measurements and the relevance to QIS would be an important contribution.
    \item Undergraduate education is becoming more important as the problems we would like to solve become more complex. New and innovative ways for teaching undergraduate quantum mechanics and QIS will be important. 
    \item Can chemical reactions be tuned or controlled by suitably entangling the reactants?
    \item It would be powerful to have some kind of local operator that enables spectroscopic (and other) measurements possible in more than one basis. 
    \item Interference is a powerful concept, are there new ways we can exploit it?
    \item A key to quantum sensing is phase retrieval. New methods development will be useful.
    \item Can quantum light push the limits of time-bandwidth product and thereby enable new pulse shaping technologies?
\end{itemize}

\mybox{{\large \textbf{Big idea:}}}{red!40}{red!10}{An experiment to measure how entanglement changes during a chemical reaction.}

\subsection{Chemistry that is governed by or optimized by QIS and opportunities presented by chemical systems as platforms for studying QIS}

\subsubsection{Executive summary}

Chemistry is defined by molecules and their complexity. Further, the properties of molecules allow them to reactant and produce new molecules. Owing to the many degrees of freedom of molecules and their encounter complexes, they interact in complex ways, and by various mechanisms. Therefore, we can think about chemical-oriented QIS as research that (i) exploits the structural control we have over molecules, or (ii) studies how quantum correlations evolve during chemical change, or (iii) uses the tools of chemistry to exhibit carefully designed molecular systems that can test or reveal fundamental aspects of QIS. 

\begin{figure}[!ht]
\centering
\includegraphics[width=1.0\linewidth]{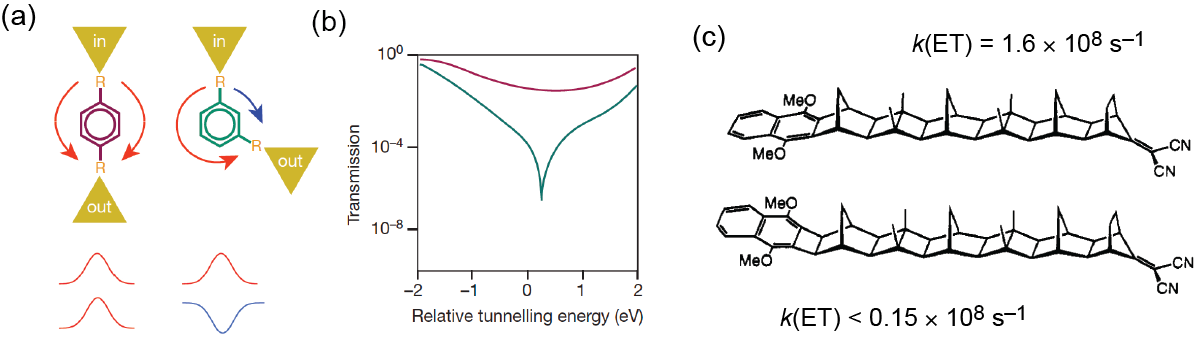}
\caption{\label{fig:3r}(a) Conceptual illustration of the electron wavefunction amplitude for two alternative transmission  paths through the molecule. R = thianoethyne. The gold triangles represent the electrical contacts of the molecular transport junction apparatus. (b) Current transmission curves predicted for the para-configuration molecule (red) and the meta-configuration molecule (green). Reproduced from \cite{scholes2017}. Original data are from \cite{solomon2012}. (c) Through bond electron transfer is diminished by a cis-linkage \cite{oevering1987, paddon1994}.}
\end{figure}

Interference is a well-known signature of phase correlations between waves or wave-like amplitudes. It can thus serve to suggest the possibility of quantum correlations in some cases. Given its prominence and relative ease of detection, it is likely that detection of interference effects will continue to yield interesting results from chemical systems. We show two examples in Fig. \ref{fig:3r} to illustrate important prior studies. Molecular transport junctions have provided fascinating examples of pathway interference\cite{solomon2012}. A highly simplified view is that when the amplitude for two pathways for carrier transport through the molecule line up in phase, the first example in Fig. \ref{fig:3r}a, we see transmission of the charge carriers through the molecule. However, when the pathway amplitudes line up out of phase, constructive interference diminishes the transmission through the molecule, by about six orders of magnitude in this example. 

A second example where interference effects are prominent, as are substantive perturbations caused by chemical structure, is through-bond electron transfer \cite{oevering1987, paddon1994}. In Fig \ref{fig:3r}c we show an example where electron transfer is mediated over 12 sigma bonds by the through-bond mechanism. The matrix element for through-bond electron transfer is very sensitive to the orbital overlaps along the path, which can be perturbed by adding a cis linkage in the norbornyl bridge. Notive how this structural change suppresses the electron transfer rate by an order or magnitude. Chemical control of this kind is a powerful tool that could be exploited creatively in QIS relevant research.

Localized or separated electron spins in molecules can be harnessed to promote chemical reactions and those spins can be controlled using magnetic fields and microwave pulses. This area of spin chemistry has escalated in recent years \cite{olshansky2019, rugg2019, wasielewski2020} and offers a clear connection with QIS. Future work might imagine how entanglement of spins be used to enable ‘quantum-concerted’ reactivity or ask questions like: can a two-step reaction be more specifically guided to a product using techniques like teleportation that can direct a sequence of spin correlations among electrons? 

We also should ask, what degrees of freedom beyond spin can be exploited for QIS? Can new non-classical states or properties be exhibited because of structure or organization of molecular systems, or the way those systems interact? One specific suggestion is to think about how the coupling of electronic and nuclear (vibrational) degrees of freedom—that is, vibronic coupling—can be used as a tool. In essence, we are entangling the electronic subsystem to structure, and structure is what changes during a chemical reaction. Is there are way, therefore, of gaining new insights into the interplay of the electronic states and reaction coordinate by watching their entanglement evolve?

Similarly, in some ways, can we work out how suitably designed molecules can serve as test-beds for studying entanglement dilution and distillation? Entanglement distillation \cite{miller2024} allows mixed quantum states to be purified into, for instance singlet states. Entanglement dilution is the reverse process, where the singlets are converted into more mixed quantum states. 

\begin{figure}[!ht]
\centering
\includegraphics[width=1.0\linewidth]{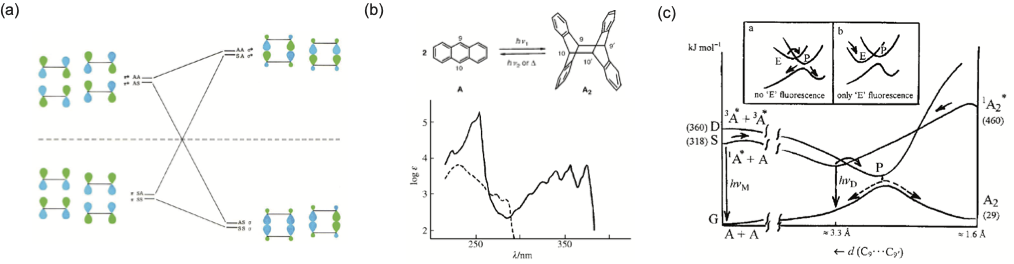}
\caption{\label{fig:4r}The Woodward-Hoffmann rules for conservation of orbital symmetry \cite{woodward1969} (a) The orbital correlation diagram for the formation of cyclobutane from ethylene. (b) Example of the photocycloadditon of anthracene (solid line absorption spectrum) to form dianthracene (dashed line absorption spectrum). (b) Example of the Woodward-Hoffmann rules in the photochemistry of anthracene photocycloadditon. The doubly-excited state (D) shares the same symmetry as the ground state product. Therefore, excitation of the anthracene singlet state is followed by a curve crossing to D, then partitioning into ground state reactant and product at the pericyclic minimum, an avoided crossing caused by the intersection of states with different symmetry.}
\end{figure}

Another direction that might be interesting is to extend the famous concept of conservation of orbital symmetry during chemical reactions, known as the Woodward-Hoffmann rules\cite{woodward1969}, Fig. \ref{fig:4r}. The essence of this theory is that pathways on potential energy surfaces connect reactants and products that have the same orbital symmetry. Conversely, the theory predicts how changing the orbital symmetry from reactant to product produces an activation barrier along the reaction coordinate. These pathways and barriers can be explained in a state correlation diagram, like that shown in Fig. \ref{fig:4r}a. The question relevant to QIS is whether we can extend this theory to consider changes in entanglement during a chemical reaction? That, in turn, would show how chemical change can be used to change entanglement.

Photochemical systems can offer excellent models for studying these kinds of reactions that are governed by orbital symmetry, Fig. \ref{fig:4r}b,c. Photocycloaddition reactions are well studied \cite{bouas2001, desvergne1983}. For example, anthracene photocycloadditon can be studied and compared with the reverse reaction produced by photoexcitation of dianthracene. In Fig. \ref{fig:4r}c we show a schematic picture of the reaction surfaces that have been inferred from experiments. Notice that the doubly excited state—the triplet pair state of the anthracene dimer—has the same symmetry as the ground state of the photodimer. 

\subsubsection{Bullet point summary of the discussion}

\begin{itemize}
    \item How long-lived should quantum effects be in order to be ‘useful’?
    \item Complex spin systems could be of interest in chemical materials, including frustrated spins and spin liquids. Here, interactions of the spins with the environment degrees of freedom becomes prominent. 
    \item Radical pair states and the way they imprint spin polarization on nuclei would be interesting to study further. They might also lead to new ideas for quantum sensors. 
    \item What is the minimum measurement protocol needed to detect quantum processes or quantum function?
\end{itemize}

\mybox{{\large \textbf{Big idea:}}}{red!40}{red!10}{Design molecular systems that allow control over, and study of, entanglement dilution and distillation.}

\subsection{Fundamental studies of quantum correlations enabled by chemical/molecular systems}

\subsubsection{Executive summary}

We think the key verb here is “to enable”. Chemical systems are incredibly diverse in structure—on multiple scales—and composition. Therefore, the variety of properties is mind-boggling. How can we exploit this platform as a way to compare or engineer quantum correlations? We might also ask: what (new) ways can non-classical correlations be exhibited and lead to new function, phenomena, states, or dynamics in fundamental chemical systems? In other words, the diversity of chemical systems may allow many-particle states with richer structures than states in simple model systems like a pair of atoms. As part of this program of work we need a formalism to assess function and correlations in more complex structures. How can the methods of quantum resource theories help clarify complex quantum processes in chemical systems? 

An example where the QIS conceptual framework can give new insights into the complex and interesting quantum states in chemical systems is recent research characterizing electronic structure and chemical bonding in molecular systems from the perspective of orbital entanglement \cite{ding2022}. Besides quantifying the quantum correlations associated to the electronic states of molecular systems, this work suggests investigation of orbital entanglement can help generalize valence bond theory beyond its current state. 

Chemical design might allow different kinds of measurements, for example by using supramolecular systems comprising strongly coupled, but spectrally distinct, subunits. As an example, can we encode correlation in three-dimensional structures like covalent organic frameworks, then use this as a three-dimensional quantum memory, addressable by two-photon absorption, for instance? Can we imagine experiments performed on supramolecular systems that detect the evolution of entanglement or coherence associated with excited state dynamics?

Another intrinsic aspect of molecular systems is their complex level structure. In particular, the interplay of electronic states and vibrational states is well studied in the context of spectroscopy. Turning that lens to quantum correlations arising from this interplay of electronic and vibrational states would be fundamentally interesting.  

\begin{figure}[!ht]
\centering
\includegraphics[width=1.0\linewidth]{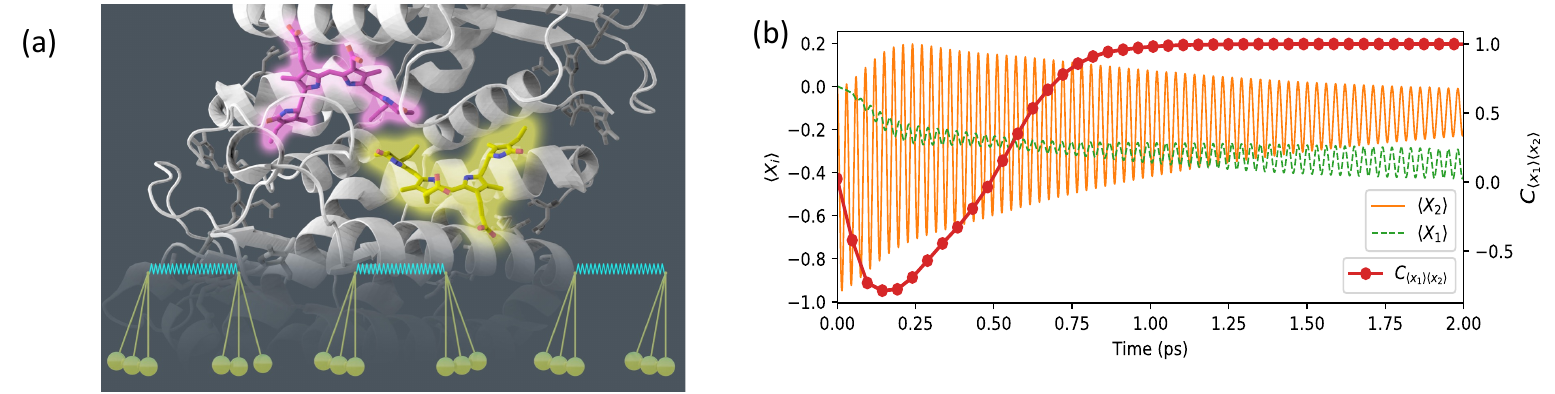}
\caption{\label{fig:sync} Transient quantum phase synchronization of local molecular vibrations mediating photoexcitation dynamics in weakly coupled heterodimers as predicted in \cite{siwiak2019} and experimentally investigated in \cite{zhu2024}. (a) Illustration of the central weakly interacting chromophore pair in the light-harvesting protein phycoerythin 545 investigated in \cite{collini2010}. The pendulum illustrate local vibrations which are involved in energy transfer. (b) The associated symmetric collective vibrational motion undergoes slow dissipation leading to positive quantum phase synchronisation of local vibrational displacements ($C_{<X_1><X_2>}\sim 1$) in the picosecond time regime. Figure reproduced from \cite{siwiak2019, siwiak2020}.}
\end{figure}

Vibronic coupling in the excited states of molecular systems can give rise to a variety of rich and interesting quantum phenomena. For example, it has been predicted that some molecular motions can undergo transient quantum phase synchronization mediated by exciton-vibration coupling while excitation energy transfer takes place \cite{siwiak2019,siwiak2020}. This is a quantum phenomenon in which a collective vibrational motion undergoes slow dissipation due to an effective decoupling from the environment via exciton-vibration coupling (see Fig. \ref{fig:sync}). Experimental evidence of this quantum phenomena has  recently been obtained via 2D spectroscopy and measurements of the dynamical Stokes shift  in an allophycocyanin complex binding  weakly electronically coupled heterodimers \cite{zhu2024}. 

Overall vibronic coupling in excitonic systems  of coupled chromophores offer a rich scenario to investigate a variety of quantum phenomena such as entanglement relations between the polarization of emitted photons and the vibrational state, in an exciton-vibrational system \cite{meskers2023,mckemmish2015}.

Chemical systems may also provide important platforms to investigate new forms of quantum-state engineering driven by dissipation where coupling to phonon environments can lead a system to particular quantum states. \cite{harrington2022engineered}.

\subsubsection{Bullet point summary of the discussion}

\begin{itemize}
    \item Some important chemical systems  have been extensively characterized and investigated from an spectroscopic viewpoint. For example, charge separated systems are well-studied and can be easily tuned (e.g., by donor-acceptor distance, redox potentials, or the reorganization energy of the solvent). Investigations of such systems from a QIS perspective can help clarifying the chemical design principles behind complex quantum processes.  
    \item The subtle correlations in supramolecular systems make development of new kinds of pump-probe techniques a priority.
    \item A key experiment is differential angular scattering as possible example of chemical studies of Bell's inequality in continuous basis, mapping phase transitions.
    \item How can we utilize topologically protected states? The concept of topological protection is not unique to quantum systems. Using the protection afforded by the topological properties, we are robust against scattering events that are unwanted for our purposes.
    \item A key priority should be standardization of common scientific language  across fields.
\end{itemize}

\mybox{{\large \textbf{Big idea:}}}{red!40}{red!10}{Identify chemical systems which allow investigating new forms of quantum-state engineering driven by dissipation }

\subsection{Measurement of quantum correlations and non-classical properties of complex systems. If we are going to look for non-classical properties, we need to be able to detect them.}

\subsubsection{Executive summary}

Exploiting the quantum nature of chemical systems requires characterization and measurement of the quantum correlations in a polyatomic system.  But measuring quantum correlations in complex systems remains a challenge, even if we were to know theoretical details of the relevant quantum state. Can experiments be devised that reveal evidence for non-classical correlations in complex chemical systems? What experiments can allow witnessing entanglement and other quantum correlations in such systems?

Entanglement is difficult to characterize experimentally as it usually requires a carefully chosen group of measurements in different basis for ensembles of identical quantum systems to allow the reconstruction of the quantum state. Such quantum state tomography is not possible or is impractical for many systems as the number of measurements needed increases very quickly with the system dimension. The challenge is then to circumvent tomography and find ways to either witnessing such quantum correlations or  estimating non-linear functionals of the quantum state described via a density matrix $\rho$ which can expose entanglement or other quantum correlations. 

A model system that was discussed extensively at the workshop is singlet fission. Singlet fission is a process where an initial singlet photoexcitation splits into a four-electron,
four-orbital singlet excited state shared across two molecules \cite{smith2013}. This fission forms a pair of excitations, with an overall singlet spin eigenstate, called  the correlated triplet pair. However, the correlated triplet pair state lies close in energy to a manifold of eight other pair states (quintets and triplets overall), enabling the initial entanglement (purity of the spin eigenstate) to be lost over a period of nanoseconds. While researchers have hypothesized this sequence, it is an open challenge to quantify the entanglement built into the triplet pair state, how it changes with time, and what causes loss of entanglement. 

Open questions include: What experiments can quantify the change in entanglement? How can we detect the entanglement spanning the molecules? Does formation of the mixed state require separation of the excitons? How can we determine how the entanglement affects the properties of the exciton pair? For example, does it change how they diffuse through the crystal by energy transfer? That is, do the excitons hop independently when entangled or do they hop as a pair state? If so, how do they separate? Lastly, how does entanglement change the properties of the excitons--can we exploit those correlation for function or as a probe somehow?

We think that one key idea to explore is multiparticle or spatial interferometry. Multiparticle interferometry has been implemented in other complex many-body systems such as in Bose-Hubbard systems.  With multiple copies of a system, quantum many-body interference allows to measure quantities that are not directly accessible in a single system such as  quadratic functions of the density matrix, like the Rényi entanglement entropy.   How can we implement similar  interferometric schemes on molecular or chemical systems of interest? 

\begin{figure}[!ht]
\centering
\includegraphics[width=0.9\linewidth]{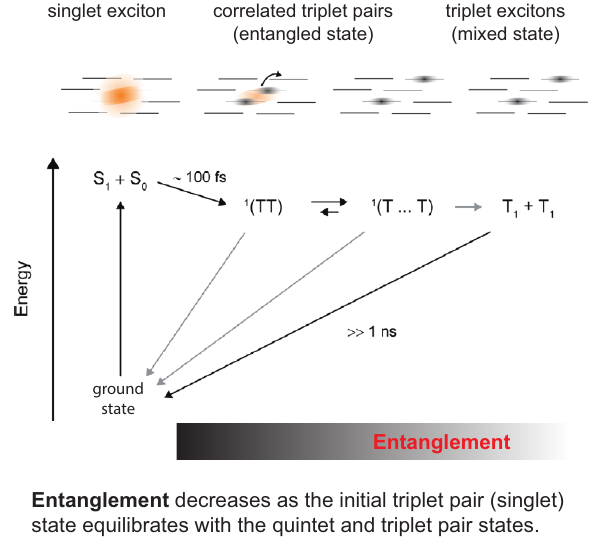}
\caption{\label{fig:singlet} Schematic of the change in entanglement during singlet fission. In this process an initial singlet photoexcitation splits into a four-electron,
four-orbital singlet excited state shared across two molecules \cite{smith2013}. This is recognized as an interesting platform to study entangled spin states across molecules.}
\end{figure}


Combination of molecular spectroscopy and quantum interferometry that exploits the correlations of entangled photons in a Hong-Ou-Mandel interferometer allows measurement of dephasing times on the order of a few hundred of femtoseconds from an organic dye in solution \cite{dorfman2021}. The question is how can such interferometric techniques be expanded to  give clear insight into the correlations within the  matter system? Spatial molecular interferometry via multidimensional high-harmonic spectroscopy that achieves attosecond and angstrom resolution appears promising \cite{koll2022}.

Another view is how extract information about the quantum correlations in matter is by using quantum optical probes. Time-resolved intensity-intensity correlations of the light emitted by a single complex system has the potential to carry out signatures on the correlations embedded in molecular system. How do the matter quantum correlations manifest in the correlations of the emitted photons? Frequency \cite{delvalle2012, holdaway2018, nation2024} and polarization-resolved \cite{sanchez2020} second-order photon correlations are emerging as tools for such studies at the single-molecule level. 

Quantum information scrambling refers to the dispersion of local information into multi-partite quantum correlations which are delocalized throughout an entire quantum system. The extent of quantum information scrambling is often assessed by studying the dynamics of out-of-time-order correlators. Theoretical studies have investigated the scrambling of quantum information in chemical reactions \cite{zhangC2024} and in molecules undergoing vibrational dynamics \cite{zhangC2022}. Such studies are highlighting important connections between the time scale of chemical reaction dynamics and the time scale of information scrambling. How can we measure information scrambling in molecular systems which are noisy and therefore will require the distinction between the scrambling and the decoherence processes affecting the quantum dynamics?

\subsubsection{Bullet point summary of the discussion}

\begin{itemize}
    \item Measures of entanglement are well-defined for bipartite systems and, in principle, a trace over all degrees of freedom, except two can be done. However, the structure of multipartite entanglement is much richer than what can be captured by any bipartite subcomponent. 
    \item Out-of-time-order correlators, which are commonly used to signal  quantum chaos in condensed matter systems, can be potentially useful in  examining  quantum correlations in the dynamics of chemical systems.  How can we implement out-of-time-order correlation measurements in complex chemical systems? Can pump-probe spectroscopy be linked to measurements of  out-of-time correlations? 
    \item It is  necessary to formulate a quantifier for why certain spectroscopic techniques are insufficient to access quantum correlations in order to unequivocally justify the need for new experimental probes beyond current spectroscopy techniques. 
    \item Time-resolved resonant inelastic x-ray scattering has been deployed to follow the photoinduced dynamics in molecular systems with attosecond and amstrong resolution. How can these measurement be exploited to witness entanglement in such systems? 
    \item Finding ways of exploiting chirality. 
\end{itemize}

\mybox{{\large \textbf{Big idea:}}}{red!40}{red!10}{Theory and experiment of out-of-time-order correlators in photo-initiated dynamics of molecular systems }

\section{Examples of research at the QIS-Chemistry intersection}

\section*{Example 1: Quantum Light Spectroscopy}

Nonlinear spectroscopy is concerned with the study of matter dynamics in response to light-matter interactions. Classical electric fields are commonly used to keep the description of the electric field state simple. Nonlinear quantum optics in contrast studies the generation of nonclassical electric field states through nonlinear light-matter interactions. For the generation of quantum light, the matter is driven far off-resonance as to not disturb the delicate quantum states and as such merely serves as a catalyst for effective interactions between different field modes.

The field of quantum spectroscopy \cite{dorfman2016, mukamel2020} aims to combine the quantum optics and the spectroscopic community by applying quantum states of light in spectroscopic techniques. This field is driven by the impressive developments in the generation and control of quantum light, which introduces additional degrees of freedom that can unravel novel and unique molecular information.
As of today, various concrete theoretical as well as experimental applications of quantum light states for spectroscopy have been developed. The merits of quantum light can be summarized in three major points, favorable intensity scaling with entangled photon sources, pathway selectivity, and joint spectra-temporal resolution beyond the Fourier limit.

A two-photon absorption process induced by two classical coherent fields scales quadratically with the two input intensities. If a single pair of entangled photons is used to induce the two-photon absorption process, the process scales linearly with the input intensities \cite{klyshko1982}. Signals with low input intensities are thus enhanced by entangled photon sources, which can be used to study photo-sensitive samples without damaging them. The experimental demonstration of entangled pair two-photon absorption has gained a lot of interest in recent years as experimental \cite{lee2006, hickam2022single} and theoretical \cite{raymer2021} estimates of the cross sections of these processes differ widely. Another interesting question concerns the cross-over between linear and quadratic scaling \cite{raymer2022,drago2022} when the number of entangled photon pairs hitting the sample is increased. 

In classical spectroscopy, only the generated field is actually measured and excitation pathways can be isolated using different phase matching configurations. Quantum light is sensitive to back action introduced by the nonlinear response. Monitoring a higher number of interacting field modes (instead of only the generated one) in terms of coincidence measurements, combined with interferometric techniques has demonstrated a higher control over the contributing pathways that goes beyond the classical control via phase matching \cite{asban2022,kizmann2023, yadalam2023}. The higher selectivity can be used to reduce the congestion of measured spectra, which is advantageous for the fitting of model Hamiltonians. This selectivity can also be used to remove unwanted background signals that can otherwise interfere with the process of interest.

Entangled photons possess unusual spectro-temporal correlations that can be used to beat the Fourier limit. This can be used to monitor molecular dynamics with high joint spectral and temporal resolution unmatched by classical sources. Entangled photons with anti-correlated frequency correlations can monitor conical intersections \cite{chen2022} and can be used to study photoelectron signals with ultrafast excited state dynamics \cite{gu2023}. Entangled photons with positive frequency correlations can enhance Raman transitions \cite{svidzinsky2021} and can also provide information about excited state dynamics \cite{fan2023entangled}.

The potential  for quantum light spectroscopy goes beyond the entangled biphoton states above mentioned. Other relevant quantum states of light that can serve as probes include single and higher photon-number Fock states, other fixed-photon number states such as Holland-Burnett states~\cite{Szczykulska2017}, squeezed states, 
photon-subtracted states~\cite{Bartley2013} and many others yet to be identified.
Furthermore, the set of relevant measurements go beyond mere intensity, correlation, or interferometric measurements. 
Indeed, the ability to identify optimal measurements that attain the fundamental spectroscopic performance allowed by the laws of quantum mechanics and statistics
via the quantum Cram\'er-Rao bound is central to quantum-enhanced sensing. 
This formalism also allows for the identification of the optimal probe states in all the relevant degrees of freedom.
The concepts of quantum optics, classical statistics and quantum estimation theory are beginning to be brought together for spectroscopy, for instance, 
using unchirped single photons~\cite{Albarelli2022}, chirped single photons~\cite{Darsheshdar24}, and entangled biphotons~\cite{Khan2024}.

The field of quantum light spectroscopy is in its infancy and is rapidly developing. It lies at the intersection of nonlinear spectroscopy and quantum optics and  a combined effort of experts of both fields will be needed to push it forward.  The main research questions of this field are very timely and tie in well with the current quantum advantage initiatives worldwide.




\section*{Example 2: Coherence and Entanglement in Reactive Collisions}

Coherence and entanglement are hallmarks and resources of quantum mechanics. How those properties evolve in chemical systems where reactions are often dynamic is a fundamental question. Chemical systems can start  in an entangled state such as electronic or nuclear spin singlet and triplet states due to the natural interaction within the system. After a reaction, are the spins still entangled and among separated products? How do we prove such entanglement? If the separated products inherited the entanglement, then the products are chemical ``Einstein-Podolsky-Rosen (EPR) pairs" that can serve as resources  for quantum communications or subsequent chemical processes. As explained in the Background (p.\,7 of this report), entanglement is peculiar. An EPR pair would be measured to have correlations of the two spins, for example, anti-aligned, regardless of the reference frame.

How general would such observations be among diverse chemical systems? And what about other molecular degrees of freedom besides spins? An affirmative answer would be the first step to stretch our imagination regarding what might be possible, maybe even connecting to applications relevant to biology. 

Recent advances in atomic, molecular, and optical (AMO) tools provided a new avenue to examine the role of coherence and entanglement in certain chemical reactions and chemical systems.  The signature of these AMO systems, although sometimes of exotic chemical species, is the high level of quantum control that can be achieved on the system (e.g. preparing the system in a single quantum state), the tuning of collision energy in the ultralow temperature range below even the energy scale of the nuclear spin, the precise tools that can be used to tease out important quantum correlation that would otherwise be averaged out in ensemble measurements.

\begin{figure}[!ht]
\centering
\includegraphics[width=1.0\linewidth]{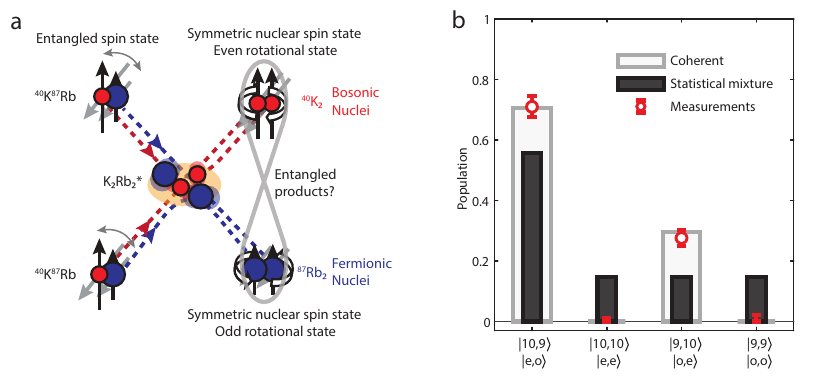}
    \caption{\label{fig:KRbreaction}(a) Atom-exchange reaction used to redistribute entanglement (b) Measured population distribution consistent with coherence is preserved throughout the reaction.}
\end{figure}

In a recent paper \cite{LiuZhu2024}, the authors examine the role of coherence and entanglement in the gas-phase atom exchange reaction of $2\mathrm{KRb} \rightarrow \mathrm{K_2} + \mathrm{Rb_2}$ (Fig.~\ref{fig:KRbreaction}(a)). Because coherence and entanglement are expected to be fragile, this work focuses the nuclear spin degree of freedom where environmental perturbation is less of a concern and the low-density, ultracold nature of the study allows researchers to examine the reaction under ``textbook conditions."  The experiment starts by first identically preparing all reactants in an entangled nuclear spin state. This is a single-body eigenstate where the entanglement exists within the individual molecules. The nuclear spins involved are spin $\lvert{\mathrm{I_K} = 4}\rangle$ and $\lvert{\mathrm{I_{Rb}} = 3/2}\rangle$. However, for illustration, we call it $\frac{1}{\sqrt{2}}\left (\lvert{\mathrm{\uparrow_K\,\uparrow_{Rb}}}\rangle + \lvert{\mathrm{\downarrow_K\,\downarrow_{Rb}}}\rangle \right)$, where the K spin is entangled with the Rb spin in each molecule. After the atom exchange reaction, K and Rb are separated and paired up with another K and Rb individually. If reactions simply swap atom partners while keeping the coherence of the wavefunction preserved, entanglement can be redistributed into the product molecules where the entangled pair of K-Rb now lives in two separate products. To examine whether coherence is preserved or not, it is crucial to pick out the reaction product pairs that come from the same individual reaction event, free of ensemble averaging. This is because without distinguishing the outcome of the individual symmetry channels, coherence-preserving vs. non-preserving processes would look identical. The measurement also makes use of the spin-rotation correspondence of homonuclear molecules to probe the parity of product molecule rotations as a measure of spin symmetry.  Coincidence detection is employed where the rotation quantum state of the products is probed and correlated to be from the same reaction via constraints of momentum conservation. The population accumulated result is shown in Fig.~\ref{fig:KRbreaction}(b), where no population is found in the destructively interfered (even, even) and (odd, odd) channels, indicating a high degree of coherence is preserved throughout the reaction. The result suggests that entanglement prepared within the reactants could be redistributed through the atom-exchange process. Note that the present experiment measured populations and the rigorous proof of the phase coherence of the products was not performed  and is reserved for future work. 

Nuclear spin coherence can be found in many chemical systems. The KRb reaction was probed in isolation, and therefore the fundamental finding is likely generalizable to other chemical species in solid and condensed phases. The tool developed here devises a way to probe coherence and entanglement and possibly in other molecular degrees of freedom. Ref.~\cite{LiuZhu2024} is a beginning. More work will need to be done and one can dream about harnessing the rich tools of chemical synthesis from the broader community for a wide variety of new applications in chemistry and quantum science.

\section*{Example 3: Ultrafast measurements}

Ultrafast science has evolved in lockstep with advances in new light-sources, enabling ever-more detailed mapping of ultrafast photoexcited processes in real time. Quantum coherence and entanglement phenomena are intrinsic in this domain and elucidation of the strongly coupled nuclear and electron dynamics constitutes a major frontier in quantum dynamics. The scientific and technological importance is broad, stretching from fundamental scientific challenges in chemistry, physics, and materials science to technologies for energy, environment, health, and communications.
 
Ultrafast experiments live in a striking symbiosis with theory and simulations. However, propagation of the many-body nuclear-electronic wavefunction requires a challenging combination of advanced electronic structure theory and dynamics methods \cite{lindh2020}, with the excited state electronic structure calculations forming a major bottleneck \cite{janos2023}. Ideas from QIS offer avenues towards more efficient and accurate treatment of electron correlation and automated selection of active space orbitals \cite{barcza2011, stein2016, ding2023}, with concepts such quantum mutual information emerging as a major descriptor of electronic structure \cite{barcza2011, krumnow2016, stein2016}.
 
Furthermore, entropy measures utilised within QIS may delineate the interplay between covariance, correlation, and entanglement in nuclear-electronic dynamics, and provide a powerful tool for revealing higher-order correlations \cite{schurger2023}. New methods that explicitly account for fermionic or bosonic symmetries could become important \cite{alon2008}, and modern versions of Glauber's coherent states \cite{kirrander2011, shalashilin2018} have the potential to elegantly accommodate entanglement between photons, electrons and nuclei. The latter hold particular promise in the realm polaritonic chemistry, which exploits the strong coupling between molecules and light-modes in plasmonic nanocavities and optical cavities to control photochemical and photophysical processes \cite{galego2016, mukherjee2023} and to achieve efficient light-harvesting \cite{bujalance2021}.
 
The challenge of interpreting ultrafast measurements can be cast as an inverse problem \cite{acheson2023}. Here, the goals of QIS overlap significantly with quantum tomography, which aims to reconstitute the density matrix for a complete characterization of a quantum system. Quantum state tomography has been achieved in a limited number of ultrafast measurements \cite{skovsen2003, bourassin2020, zhang2021}, with molecular frame quantum tomography of a dynamical polyatomic system recently demonstrated \cite{morrigan2023}. Reconstruction is limited by the number of observations, incomplete sets of observables, or low fidelity. Maximum-entropy based approaches developed in the context of QIS provide a robust method for density matrix reconstruction \cite{gupta2021}. The ultrafast community, in turn, has begun to explore multimodal measurements that project different aspects of the quantum state onto complementary observables, for instance by combining spectroscopic and scattering techniques. These developments will benefit quantum technologies by providing sensitive diagnostics of quantum states.
 
The application of quantum information concepts to ultrafast dynamics is likely to provide new routes for controlling physical and chemical processes. Recent advances in attosecond pulse-generation offer avenues towards measurements of the transfer (or loss) of coherence as excited electrons couple to nuclear dynamics \cite{kretschmar2024}. This provides an exciting opportunity to probe fundamental correlations in molecules, and tackle long-standing questions regarding the transition from charge migration to charge transfer. In photoionization, theory predicts that chirped ionisation pulses may control the degree of entanglement between the bipartite subsystems \cite{vrakking2022} and that pulse sequences can be designed to maximise the degree of coherence in the molecular ion \cite{dey2022}. The former concept has recently been demonstrated experimentally \cite{koll2022}, illustrating the inherent trade-off between coherence and entanglement \cite{ruberti2022}.
 
An exciting aspect of time-resolved spectroscopies is that they can be exploited as a new platform for quantum information processing. The strong correlations between electronic and nuclear degrees of freedom in photoionization, gives rise to coupling and potential entanglement between electronic, vibrational, rotational, and, potentially, spin degrees of freedom. This has non-trivial dynamical and spectroscopic consequences that provide opportunities for technological exploitation such as information processing. Ideas are emerging for information processing in the attosecond regime, for instance via interferences between different electron trajectories in high harmonic generation (HHG) and the entanglement between the generated XUV photons and residual infrared photons \cite{lewenstein2024}. Others have proposed that nonlinear spectroscopy in assemblies of nanosized colloidal quantum dots could constitute an effective platform for quantum information processing, programmable via Lie algebras \cite{remacle2023}, with recent comparisons between simulated and measured time-frequency polarization maps giving credence to this concept \cite{hamilton2023}.
 
In summary, there are strong and tangible mutual benefits between QIS and ultrafast science. Concepts from QIS will aid experimental capabilities in the ultrafast regime, provide new tools for the interpretation of measurements, and advance computational methods. Reciprocally, the same advances will benefit quantum technologies via better simulation methods and better quantum-state diagnostics. Perhaps most intriguing are the emerging concepts for quantum information processing based on various forms of spectroscopic and ultrafast measurements.

\section*{Example 4: Photon correlation  measurements}

Light-matter interactions are one of the most accessible ways to create and read-out state behaviour in materials. Conventional measurements in the optical domain – for example time-integrated or transient photon absorption or emission – reveal the temporal and/or energetic population of these states, but typically say little about any role of quantum phenomena in how those states interact with each other. To understand, enhance or exploit any quantum behaviour in materials, methods to reliably measure it are needed.

The most successful technique developed to-date is based on an optical analogue of the Hahn Echo. These experiments, 2D photon Echo (2DPE) spectroscopy \cite{mukamel2000, tian2003, pisliakov2006, engel2007, collini2010}, rely on using a laser pulse to first create an excited state, then deliver subsequent laser pulses to interact with that state and readout the photon echo. 2DPE is an ensemble technique and has the advantage of being able to potentially detect excitonic coherences within an inhomogeneously broadened ensemble because it is using the interaction of two identical pump laser pulses to enable something incoherent (many different excited states measured at once) to become coherent again (all excited states are subject to the same 2\textsuperscript{nd} pump pulse) and be read out. The 2DPE measurements are an inherently perturbing experiment, as they rely on deliberately inducing a new signal (the echo) as a readout for state coupling and coherent effects. This is a requirement when one is measuring ensemble samples where many thousands (or millions) of molecules are simultaneously excited, as is almost always the case in traditional spectroscopic techniques. It is appealing to find new methodologies that do not rely on perturbing the system, but instead use the excited state itself as a readout of its behaviour.

To use output from the system for evidence of quantum behaviour, single objects are an attractive regime to consider, as they avoid the inherent randomisation averaging that comes when working with ensembles. These single objects can be in the form of aggregated particles, quantum dots, single polymer chains or single molecules. The quantum nature of these is supported by the observation of single photon emission, which is evidence of a two-level quantum system \cite{michler2000}. Here the second order emission intensity cross-correlation ($g^{(2)}$) is measured using two detectors in a Hanbury Brown and Twiss geometry \cite{brown1956} and photon antibunching ($g^{(2)} \sim 0$) indicates that the system is a single photon emitter, a conclusion solely drawn from photons emitted by the object itself. These g\textsuperscript{(2)} measurements are potentially powerful descriptors of state behaviour in single objects, and reveal information beyond simple populations, up to and including quantum phenomena. Short examples are detailed here for consideration. 

In contrast to single photon emission, entangled photon \textit{pairs} are attractive to utilise in quantum information science. These can enable secure communication, ghost imaging or new spectroscopic techniques, \textit{e.g.}herald photon spectroscopy. Generation of entangled photon pairs is typically achieved using spontaneous parametric downconversion of laser light, however alternative methods such as emission from quantum dot biexciton states or spontaneous four-wave mixing can be used. An alternative to achieving the entanglement of photons through their polarisation is to use their time-energy interaction. Strong resonant excitation of a two-level system can lead to the generation of a Mollow triplet, so-called due to the formation of two sidebands at higher and lower energy of the main transition. Under such strong excitation, scattering of photons as well as fluorescence is present. Overall, photon antibunching will be observed ($g^{(2)} < 1$), however, more information can be extracted if spectral filtering of photons on both detectors is applied \cite{cohen1979, delvalle2012, holdaway2018}. One example of this is when measuring epitaxially grown InAs quantum dots at 5 K \cite{peiris2017}. Here by observing $g^{(2)}$ values across an array of filter energies on both detectors a map can be constructed. This evidences that despite the wavelength integrated $g^{(2)}$ being $< 1$, there exists regions in the $g^{(2)}$ map that show values significantly above 1, with the map revealing that not only cross-sideband photon bunching, but also interactions between each of the peaks (band, sideband), producing a fingerprint of state interactions. Furthermore, it was found that the filtered bunching peaks produce photon pairs close to violating Bell’s inequality for entanglement, indicating these methods could be used as a generation source in quantum optical applications. 

The concept of energy and time filtering of photons in the generation of correlations can be expanded beyond grown quantum dots to instead explore single molecules. Here the experimental conditions are more demanding, thus theoretical exploration of these has so-far been the primary way that they have been understood. Firstly, considering a single two-level dye molecule in a single molecule measurement \cite{bel2009}, the bandwidth of the energy filtering is found to play an important role in the observed correlations. Correlations between fluorescence photons at significantly different energies is primarily responsible for antibunching, while fine structure exists linked to intersideband interactions. These results for a simple two-level system follow the same behaviour as described above for the InAs quantum dot, however, these maps can become very rich when considering more complex situations, \textit{e.g.} two coupled emitters \cite{darsheshdar2021, nation2024}. Finally, placement of a molecule in an optical cavity can create hybridised states, exciton polaritons. Measurement of energy filtered correlations could then be used to observe interactions between dressed Jaynes–Cummings ladder polariton states \cite{dorfman2018}. Additionally, it has been suggested that this information could enable coherent control of the polariton via feedback into coupling time profiles. 

In summary, it is clear that energy and time filtered photon correlation measurements can enable access to rich information on state interactions. Consideration must thus be given on how these can be routinely achieved experimentally, where photon rates are low and the measurement timescales demanding.

\section*{Example 5: Molecular spin-based qubits}

Within the context of quantum resource theories \cite{streltsov2017}, quantum coherence can serve as a resource state via interference phenomena or other non-classical correlations. Such resources are important for sensors and detectors, and provide the possibility for chemical manipulation of quantum correlations via manipulation of coherent states. Sensing requires the coupling of a responsive element, which can be bistable, to a reporter, which can be classical or quantum in nature. The gating of molecular structure through photoinduced excited states provides an interesting paradigm for sensing on the molecular level \cite{wasielewski2020}.  Photoinduced cis-trans isomerization in the rhodopsin retinal chromophore involves wavepacket evolution on the fs timescale and strong vibronic coupling that results in coherent oscillations even after photoisomerization is complete \cite{chuang2022, shigaev2020, videla2018}. Light-induced magnetoreception, an important example of biological magnetic field detection, operates via changes in the population of singlet-triplet states in charge-separated radical pairs formed from photoexcitation in the cryptochrome family of enzymes \cite{hiscock2016}. 

By electronic coupling optically bistable molecular states, via molecular switches, to a two-state quantum spin system, such as a molecular spin $S=1/2$ state, with \textit{m\textsubscript{s}} $\pm 1/2$ microstates, a four state system is in principle created, in which the population of each of these four states can be manipulated by environment. Within the $S=1/2$ spin system, resonant microwave excitation leads to a coherent state at time = 0 that gives rise to a Rabi oscillation with a frequency on the timescale of 10s of ns for most molecular systems. Manipulation of the Rabi frequency, which provides a basis for the spin projection operator, leads to a possible strategy for sensing via changes in coherence. Such a strategy was first proposed for optically-gated spin qubits \cite{paquette2018, jenkins2018} in which photoisomerization in a metal-ligand system leads to a change in the driving force for a charge transfer-induced spin transition, leading to significant changes in spin state, oxidation state, and spin relaxation dynamics at the paramagnetic metal center. Significant changes in the coherence time and energy gap between \textit{m\textsubscript{s}} $\pm 1/2$ quantum states (g-value) are observed with photoisomerization, suggesting a method for probing the effects of structure, spin density distribution, and metal-ligand covalency on the nature of coherent states in molecular spin-based qubits. 

The in-situ manipulation of coherence in spin systems for sensing protocols requires modulation of contributions form through-bond magnetic exchange, through-space dipolar interactions, and electron-nuclear hyperfine and super-hyperfine interactions in response to an external stimulus. The stronger the spin-spin interactions, the shorter the coherence time, posing challenges for entangled molecular spin systems, in which one predicts a decrease in coherence time with exchange. Photoisomerization of a coordinated ligand however results in a change in metal-ligand covalency, which in turn modifies the spin density distribution in the ligand, and in turn, hyperfine and superhyperfine couplings. The change in spin-spin couplings results in changes in the coherence time as well as Rabi frequency, allowing fundamental insight into the structural changes that control quantum spin-state dynamics in molecular systems in which the bath, geometry, local environment, and phonon modes of the bath are held constant \cite{ghosh2024}. These studies highlight metal-ligand covalency as being critical to controlling subtle changes in coherence, and allow determination of possible design strategies for molecular quantum sensors for QIS technologies.

\section{Open research questions}

\subsection{Executive summary}

A defining characteristic of molecules is that they are intrinsically complex quantum systems with many degrees of freedom, exhibiting remarkable variety. Molecules, therefore, have different quantum properties than atomic systems, including solids. Our aim should then be to leverage these distinctions rather than trying to solely translate paradigms from the QIS field to chemistry. This is where transformative advancements lie. 

Chemistry brings to the field the prospect of modifying molecules in countless ways. They can be functionalized, joined together, unhinged, rigidified, and so on. Exploiting this core ability will enable unprecedented  comparisons of molecular quantum phenomena.

Each molecule is a rich complex quantum system, exhibiting hierarchical entanglement of electronic, vibrational, and even rotational coordinates. This suggests that studies of how the Born-Oppenheimer approximation weakens are of relevance. Examples include studies of vibronic coupling in excited state dynamics and ambitious experiments, like TRUE-CARS (Transient Redistribution of Ultrafast Electronic Coherences in Attosecond Raman Signals), that have been proposed  to map out dynamics through conical  intersections \cite{kowalewski2015}. We may also consider further studies beyond potential energy surfaces, for example, to properties of non-adiabatic coupling manifolds that arise from their topology.

The downside of the many internal degrees of freedom of molecules, and also the way they interact strongly with their environment, is that non trivial quantum states tend to be fragile—they are subject to rapid decoherence. Challenges for the field therefore include: How to characterize sources of decoherence? Can decoherence be mitigated? Can we live with strong decoherence in one basis by focusing on another basis? How can we exploit dissipation processes to lead a chemical system to a targeted quantum state? 

One possible way forward is to imagine a hybrid of quantum and classical worlds. In the classical world, weakly coupled oscillator systems can form spontaneous collective sates: if started out of phase, they can synchronize spontaneously. The larger the system of coupled oscillators, the more robust such syncrhonized states tends to be. Synchronization abounds in Nature and in engineered systems. It coerces complex systems into unified and remarkable collective functions. It is evident from many examples in the natural world—from the beating of a heart to plumes of flashing fireflies. It may be possible to find a way to use classical synchronization to stabilize a system so that it becomes a more robust host for quantum states. 

On the other hand, ultrafast transient spontaneous synchronization of molecular motions can also happen during photoexcited dynamics despite strong coupling to environments, indicating that molecules have interaction mechanisms which lead to effective decoupling of collective quantum modes from the environment.

Electrons, spins, and electron-electron interactions are the "fundamental particles and forces” in chemistry. It makes sense that they serve as resources for QIS.  There has already been extensive attention paid to spin states as prototypical resources for entanglement. Ongoing  research will investigate the dynamics of these systems after photoexcitation. Relevant systems includes spin systems formed by singlet fission, excited state  dynamics of radicals, biradicals, etc, and weakly coupled spins produced by charge separation. Investigations of multi-spin systems will enable elucidation of quantum correlations beyond the well-studied Bell states and may yield model systems for demonstrating measures of entanglement and obtaining new  viewpoints on the many-body problem. Investigations of the ultrafast dynamics of populations will be complemented by studies of the evolution of spin densities. Molecular synthesis will be exploited to control the dynamical evolution of quantum correlations evidenced in spin densities during photochemistry, reactions, or photophysical processes.

The fundamental interactions between electrons may also receive attention.  In particular, the exchange interaction is the quantum mechanical correction to the classical Coulomb interaction between charge densities. Exchange accounts for the way like-spin electrons  naturally avoid each  other because of the  Pauli  principle, thus, ameliorating the blow-up of the Coulomb interaction when electrons are closely separated. The exchange interaction is implicated in stabilization of chemical bonds as well as the splitting in the spectrum of singlet and states from triplet states. The exchange interaction can be used to indicate size and shape of wave functions. Future work may use such probes to map dynamics of charge density or states.  

A key challenge for the field is how to measure correlations that are unique to multipartite quantum states, including entanglement. This is an important challenge because it would allow clear answers to  questions about "quantum advantage" in the properties of molecular systems. However, even with access to the density matrix of a complex quantum system (beyond Bell states) it is not clear that a state can always be characterized as separable or non-separable. Nevertheless, design of suitable witnesses can provide useful "litmus tests" of quantum correlations.

In sum, a key conclusion of the workshop discussions was that leveraging the rich properties of molecules for QIS-relevant research is an open direction for future work. It is envisioned that the chemist's expertise at constructing and editing molecules on a range of scales will  provide model systems where thoughtfully designed experiments can reveal influential, fundamental insights into QIS and the quantum world at the molecular scale. Joint theory and experiment research will be essential in achieving this goal. These advances will enable chemistry, as a field, to advance new directions for quantum science.

\begin{mdframed}[style=MyFrame, nobreak=true]
\begin{Large}
    \textbf{Box 2: Molecular QIS wish list—Six ideas}
\end{Large}
\\

\begin{enumerate}
    \item Can we find molecular or chemical versions of quantum simulators? Perhaps first steps will involve constructing and working out ways to address three-dimensional arrays based on covalent organic frameworks.
    \item  Molecules are unequivocally quantum mechanical entities. But the way we think about chemical reactions and, often, how we model them is largely classical. Where is the quantum-classical “dividing line” for chemical reactions and what does it mean?
    \item Are there ways we can use weak measurement to probe the quantum mechanical basis for chemical dynamics? Can such a method find evidence for quantum correlations in chemical systems? 
    \item Demonstrate that entangling reactants in different ways can influence a chemical reaction 
    \item How can we "live with decorehence"? Devise an abstraction of our usual view of chemistry where intermolecular entanglement is more robust.
    \item Develop a “toolbox” of witness protocols that aid detection of non-classical correlations in complex molecular systems.
\end{enumerate}
\end{mdframed}

\section{Broader Impacts}

This workshop aimed to specify substantive ways that chemical sciences can create new opportunities for knowledge and, perhaps, technologies that use or enable quantum resources. As such, we anticipate that the outcomes, summarized in this report, will engage and inspire scientists of all career stages. We hope that the report will inspire new research programs, new research directions, and find new opportunities for the methods, experiments, theories and facilities that have been developed to address research questions in chemistry. 

It is foreseen that, in turn, new and interesting discoveries in chemical sciences will  impact education and public perception of chemistry. We anticipate that new research and teaching on the theme of the report will particularly advance the training of high qualified personnel in the intersection between chemistry, physics and QIS —training them to think differently about where opportunities for new science may be explored. 

One of the key goals of the \href{https://www.gov.uk/government/publications/national-quantum-strategy}{UK National Quantum Strategy}  is  "\textit{to ensure the UK is home to world-leading quantum science and engineering, growing UK knowledge and skills."}  Research programs at the intersection between chemical sciences, physics and QIS  will be essential for  building  innovative scientific communities that will advance  quantum science beyond what we can envision today. 

The UKRI-EPSRC are committed to advance fundamental science for new quantum technologies. Indeed,  \href{https://www.ukri.org/what-we-do/browse-our-areas-of-investment-and-support/physical-sciences-theme/}{Quantum Physics for New Quantum Technologies} is one of the Physics Grand challenges within the Physical sciences theme of UKRI-EPSRC. The workshop and the associated report directly align with the EPSRC aim to bring scientific communities to work together to accelerate progress towards major scientific breakthroughs in this area by exploring the research opportunities and new horizons at the intersection between chemical sciences, physics and QIS.

Similarly,  according to \href{https://www.nsf.gov/mps/quantum/quantum_research_at_nsf.jsp#Workshops}{Quantum research at NFS}, the NSF is committed to continuing to foster quantum-based research in the following three ways:

\begin{itemize}

\item \textbf{Advancing Quantum Frontiers}: Frontier knowledge generated through NSF-supported discoveries will open new vistas and opportunities in the quantum arena, such as new materials, circuits, and algorithms that enable novel quantum and post-quantum applications including artificial photosynthesis, highly sensitive radiation detectors, and many others not currently foreseen.

\item \textbf{Multidisciplinary Collaboration}: NSF will capitalize on the full breadth of scientific and engineering areas that it funds to bring together researchers from multiple disciplines to address the fundamental science and engineering questions that will accelerate progress in all areas of quantum applications, from sensing to communication to computing to simulation.

\item \textbf{Workforce Development}: Through its support for research and education at universities, NSF investments will build capacity by training the workforce that is essential to progress and commercialization in this rapidly expanding field of emerging technology.

\end{itemize}

NSF's investments are aligned with the National Quantum Initiative and address the policy goals expressed in the National Strategic Overview for Quantum Information Science. NSF's continued support is expressed through investments in the core NSF disciplines as well as investments in specialized activities having specific targets, many of which overlap with NSF's Quantum Leap Big Idea.

This report  will inspire in each of these areas. For example, by articulating research needs and key opportunities, we motivate new research proposals and plans that will advance quantum frontiers from the chemical perspective. The workshop outcomes will undoubtedly also motivate how and why interdisciplinary team science will play an important part of the future significant advances in QIS from a chemical perspective, while taking advantage of tools and theories from physics. We also anticipate that the workshop will identify and nucleate opportunities that leverage capabilities in the US and UK and thus build partnerships. A wider outcome of all these broader impacts will be training of interdisciplinary scientists who will be key players in the workforce of the near future. 

Another outcome of the discussions in the workshop is the need for new ways to teach quantum mechanics and introduce quantum information science in the undergraduate curriculum. A challenge is the extensive foundational material that we currently teach, leaving little time for more modern topics. The other challenge is that QIS is technical, subtle and complicated. The on-ramp for teaching even a centerpoint, like entanglement, properly is out of reach at the undergraduate level. However, are there other ways of introducing quantum information science at the undergraduate level? Ideas along this direction will be worthwhile exploring.  

\section{Acknowledgements}
We are grateful to all the workshop participants for the lively and thoughtful discussions at the workshop. This report, of course, has been greatly inspired by the ideas generated by the participants. The scribes did wonderful work documenting these discussions. We thank the NSF staff who attended the workshop. Notably, Dr Denise Caldwell, Acting Assistant Director, Directorate for Mathematical and Physical Sciences, NSF presented open remarks, and David Berkowitz, Division Director of Chemistry opened the second day. We thank Nagwan Ali (Princeton University) and Alex Balçiunas (UCL) who tirelessly took care of administrative details. We also thank Christian David Rodriguez-Camargo for helping with the pre-workshop assessment forms and the formatting of this report. This workshop was supported by awards from NSF (CHE-2403812) and UKRI-EPSRC (International Science Partnerships Fund).

\printbibliography
\end{document}